\documentclass[nofootinbib,
 amsmath,amssymb,
 aps,twocolumn,nofootinbib
]{revtex4-1}
\usepackage[parfill]{parskip}
\usepackage{graphicx}
\usepackage{amssymb}
\usepackage{amsmath}
\usepackage{mathtools}
\usepackage{color}
\usepackage{caption}
\usepackage{subcaption}

\newcommand{\be}{\begin{equation}}
\newcommand{\ee}{\end{equation}}
\newcommand{\ba}{\begin{eqnarray}}
\newcommand{\ea}{\end{eqnarray}}

\begin{document}
\title{Galaxy rotation favors prolate dark matter haloes}
\author{Adriana Bariego Quintana, Felipe J. Llanes--Estrada and Oliver Manzanilla Carretero}
\affiliation{Univ. Complutense de Madrid, dept. F\'isica Te\'orica and IPARCOS, Plaza de las Ciencias 1, 28040 Madrid, Spain.}

\begin{abstract}

The flattening rotation velocity $v(r)\to {\rm constant}$ found by Vera Rubin and collaborators and very apparent in the SPARC galaxy--rotation data coincides with Kepler's law in one less dimension. Thus, it is naturally reproduced by elongated dark matter distributions with the axis of prolateness perpendicular to the galactic plane. \\
This theoretical understanding is borne out by the detailed fits to the rotation data that we here report: for equal dark matter profile, elongated distributions provide smaller $\chi^2$ than purely spherical ones. We also propose to use the geometric mean of the individual halo ellipticities, as opposed to their arithmetic average, because $s=c/a\in (0,\infty)$ corresponds to spherical haloes for $s=1$, so that the usually reported average is skewed towards oblateness and fails to reveal the large majority of prolate haloes. Several independently coded fitting exercises concur in yielding $s<1$ for most of the database entries and the oblate exceptions are understood and classified.
This likely prolateness is of consequence for the estimated dark matter density near Earth.

\end{abstract}

\maketitle

\section{Introduction: galactic rotation}

The rotation curve $V\left( r \right)$ of a spiral galaxy is the 
average of the rotational speed of stars and gas versus their radial distance to the galactic center and is accessible by the Doppler effect~\cite{lelli,lellithesis,sparc}. 
Spiral galaxies equilibrate the centripetal acceleration of their distinct rotation with, presumably,  their gravitational field. 
Therefore, from their rotation curves $V\left(r\right)$, we should be able to extract their mass distribution.

The first measured rotation curve, of M31, dates from 1939~\cite{Babcock:1938}. Horace Babcock found higher values of the rotational speed than expected from observations, implying that the mass-to-light ratio $\Upsilon=M/L$ increased radially. This was initially attributed to interstellar extinction or the need to introduce new dynamic effects. Since the late 1950s rotation curves have been  measured with the Doppler effect on the HI  line ($21$ cm)~\cite{Hulst:1957,Volders:1959}; for long, the discrepancy between the visible mass, derived from photometry, and the dynamical mass, derived from the rotation curve, was attached to the presence of dwarf stars and intergalactic dust and gas.

The work of Vera Rubin and Kent Ford in the late 1960s and early 1970s was crucial to  understand rotation curves. Their improved accuracy led Vera Rubin~\cite{Rubin:1978} to discover the general flattening $V(r)\to {\rm constant}$ of the rotation curves at large $r$, establishing the mass discrepancy as a general rule.
Since then, rotation curves have remained a current topic of research to address the nature~\cite{Bar:2021kti} and distribution of dark matter~\cite{Loizeau:2021bum}.

The renowned observational mass discrepancy arises from multi-wavelength observations of galaxies, localising visible matter within a finite volume~\cite{Mihos:2012}, in whose interior the rotation curve should increase~\footnote{This is usually exemplified by a solid sphere of constant density $\rho=\rho_{0} \Theta\left(r-r_{0}\right)$,  with gravitational field  obtained from~Eq.(\ref{eq:gravgauss}), where the mass enclosed at radius $r$ is  $m=\frac{4\pi}{3} r^{3}$ for $r<r_{0}$ and $m=\frac{4\pi}{3} R^{3}$ for $r>r_{0}$. Thus, the gravitational field $g\propto r$ linearly increases inside the sphere} ($V'(r)>0$), but outside which 
it should  decrease ($V'(r)<0$).
Outside a spherical distribution, or any distribution at sufficient distance, $g = \frac{Gm}{r^2} $ means that 
\begin{equation}
V = \sqrt{Gm/r}\neq {\rm constant}
\end{equation}
at variance with observations;
instead of declining, measured rotation curves quite generically become independent of radius $V\sim $ const. (see Fig.\ref{fig:grafanalisis} below).

Extinction by gas and dust cannot explain the mass defect: although in the visible the apparent mass would be lower, in the infrared it would increase as it is the region of gas and dust emissions.
Thus, we need to adopt one of two hypothesis, either the existence of invisible 
dark matter, or the failure of the theoretical tools requiring  modified gravitational or dynamical theories. 

\begin{figure}
    \centerline{
    \includegraphics[width=0.4\textwidth]{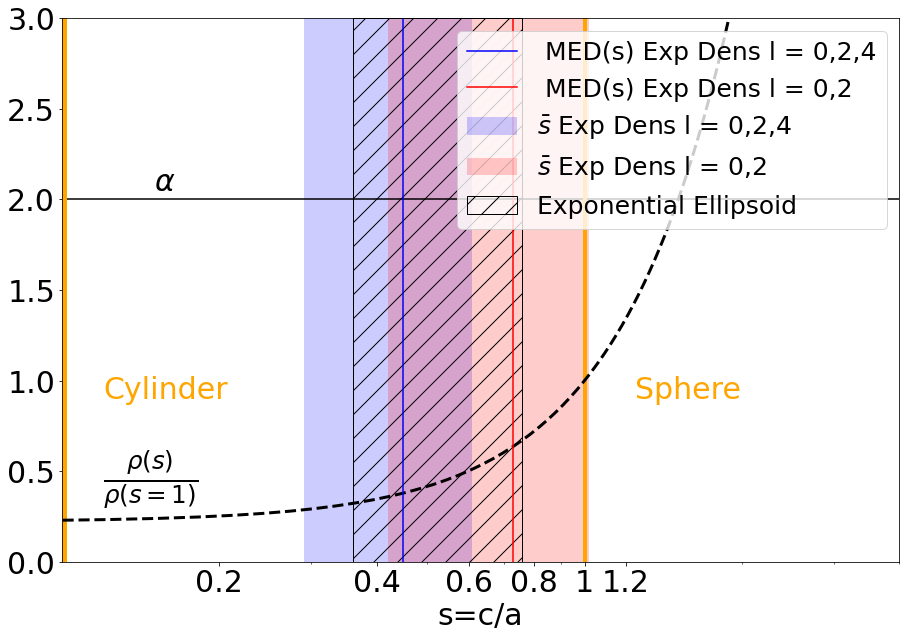}}
    \caption{\label{fig:ataglance}
    Typical DM haloes fit to rotation-curve database are clearly prolate, as indicated by the median ellipticity (two calculations at different orders of an angular expansion~\ref{secciondensidad} shown) and a nonspherical ellipsoidal model (subsec.~\ref{subsec:expellipsoid}) that can {\it a priori} be either prolate or oblate; a few outliers described below induce the uncertainty on the averages. The prolateness is in qualitative agreement with simulations (subsec.\ref{subsec:comparison}) and is true for generic dark matter profiles; however, if the exponent of the dark matter profile $\alpha = - \frac{\log \rho_{DM}}{d\log r}$ is near (the isothermal) 2, the shape cannot be decided, as there is a degeneracy between this fine-tuned spherical profile and the prolate shape. A curve detailing typical DM density in the galactic plane is also shown.}
\end{figure}

Among the solutions proposed to this problem, the approach of~\cite{Llanes-Estrada:2021hnt} is of particular interest for this work: a cylindrical source of gravitational field with mass per unit length $\lambda$
yields precisely 
\begin{equation}
V(r) = \sqrt{2G\lambda}
\end{equation}
(from the field in Eq.~(\ref{cylinderg}) below).
While that constancy of $V$ extends to the largest $r$ measured, because data covers in the end only finite  $r$, an exactly cylindrical/filamentary contribution may not be needed, and a prolate dark matter halo may suffice. In this work we report on statistical estimates, with theory fits to the database, about the optimal prolateness that observed galactic rotation curves require. Figure~\ref{fig:ataglance} shows the main result of the article: the statistical distribution of rotation curves does suggest prolate haloes.

That the analysis of galactic rotation curves can be performed with Newtonian Dynamics follows from typical velocities $\sim 100$ km/s, four orders of magnitude lower than the speed of light, and gravitational potentials  of order of the Milky Way's. With $M\sim 10^{12}M_{\odot}$, $R\sim 30$ kpc,  $\Phi\approx\frac{GM}{R}\sim 10^{11}$ m$^{2}$/s$^{2}$ and $\left|\frac{\Phi}{c^{2}}\right|\sim 10^{-5}\ll 1$. Hence, all conditions for the Newtonian limit are fulfilled and General Relativity is unnecessary. 
This allows to study the gravitational field and the rotation curves within the Newtonian framework or modifications thereof.

\newpage

The total gravitational field is the linear superposition of those of the different source components.
If there are $N$ such components, circular orbital equilibrium reads
\begin{equation}
\label{eq:vgform}
	\frac{V^{2}}{r}=-\sum^{N}_{i} g_r|_i
\end{equation}
where the radial component of the gravitational acceleration field $g_{r}$ follows from either the gravitational potential $\Phi$ as $g_{r}=-\frac{\partial \Phi}{\partial r}$ or directly from the mass density $\rho$ through Gauss' Law,
\begin{equation}
\label{eq:gausslaw}
	\int_{\partial\Omega}\vec{g}d\vec{S}=-4\pi G m\ ,
\end{equation}
With $m=\int_{\Omega}\rho dV$ interior to $\Omega$. 
Choosing the boundary surface $\partial \Omega$ as equipotential, we have of course
\begin{equation}
\label{eq:gravgauss}
	g=-\frac{4\pi G m}{S} \ .
\end{equation}
The modulus $g$ and radial component $g_{r}$ of the gravitational field coincide on the galactic plane under azimuthal symmetry around $OZ$.
Thus, the expected rotation curve $V\left(r\right)$ depends on the parameterization of the gravitational potential or the mass density (these are connected through Poisson's equation $\nabla^{2} \Phi=4\pi G\rho$). 

The puzzle of the constant rotation velocity has a natural solution~\cite{Llanes-Estrada:2021hnt} that can arise either in the dark matter or in the modified gravity scenarios, 
that of an elongated matter source. 
Indeed, in the limit of a perfect cylindrical source of linear mass density $\lambda$, the external field
\begin{equation} \label{cylinderg}
g= \frac{2G\lambda}{r}
\end{equation}
immediately yields $V(r)={\rm constant}$.
This can also be achieved by modifying gravity so that, in effect, there is one less dimension, such as in MOND --Modified Newtonian Mechanics-- (or, newly, in fractional gravity~\cite{Varieschi:2020hvp,Giusti:2020rul}).

\subsection{Summary of findings}

There have been recent attempts at fitting galactic rotation curves with nonspherical distributions; their results seem to be contradictory. While Zatrimaylov~\cite{Zatrimaylov:2020hts} seems to concur with our observation that filamentary sources offer a better fit, Loizeau and Farrar~\cite{Loizeau:2021bum}
seem to find that a disk-shaped DM component could be at play, more in agreement with vintage work by Blanco and Mercader~\cite{Rodrigo-Blanco:1996rnv}.

In this work we try to clarify the situation with systematic fits to as large a subset of the SPARC database as is possible to obtain a positive number of degrees of freedom in each situation, with several approaches. 
We first examine traditional dark matter parametrizations (with spherical geometry), MOND, and elongated geometries, against that database of spiral rotation curves, with mixed results, in which both spherical or cylindrical geometries can describe the data with a modest but sufficient parameter number to provide some flexibility. This probably explains a part of earlier discrepancies. Spherical geometries however need mass models that are similar to the isothermal one $\rho(r)\propto r^{-2}$ whereas cylindrical geometries do not need this restriction.

We then turn to a systematic multipolar expansion of the gravitational potential, and to an alternative multipolar expansion of the DM density. In both cases, with a fixed distance profile that is not the typical isothermal $\rho(r)\sim 1/r^2$, but rather arbitrary ones such as exponentials, step functions or their softened Woods-Saxon profiles, for example, we find that the rotation data prefers an elongated source. We characterize this elongation by an ellipticity variable $s:=(c/a)=(b/a)$ that speaks of a rather prolate ellipsoidal distribution.

The distribution over the galaxy population  in terms of the ellipticity is best described in logarithmic scale, 
since $s \in (0,1)$ corresponds to a prolate halo (what we find), $s=1$ to a spherical halo and $s\in(1,\infty)$ to an oblate one: the geometric mean is more reliable than the arithmetic one to avoid biasing the average to more oblate distributions.

A preliminary brief summary of our results was presented to the EPS-HEP 2021 conference~\cite{Quintana:2021ken}. This manuscript is the full documentation of the effort.

\subsection{Dark matter radial profiles and number of parameters}

Each density parameterization has a number of degrees of freedom, which is the sum of the number of free parameters and ``hidden parameters'', so called because they are prefixed free parameters without a clear physical motivation. 
For example, Navarro-Frenck-White's parameterization (NFW)~\cite{Navarro:1995}
in Eq.(\ref{eq:paramexplain}) is a specific case of Hernquist's \cite{Hernquist:1990} in Eq.(\ref{eq:paramexplain2}) for $\alpha=1$, $\beta=1$ and $\gamma=2$. 

\begin{eqnarray}
\label{eq:paramexplain}
	\rho_{\rm H}\left(r\right)&=&\frac{\rho_{0}}{\left(\frac{r}{r_0}\right)^{\alpha}\left[1+\left(\frac{r}{r_0}\right)^{\beta}\right]^{\gamma}} \rightarrow \\ \label{eq:paramexplain2}
	\rho_{\rm NFW}\left(r\right)&=&\frac{\rho_{0}}{\frac{r}{r_0}\left(1+\frac{r}{r_0}\right)^{2}}
\end{eqnarray}

These three exponents are ``hidden parameters'', as they have no obvious physical motivation. The degrees of freedom for the NFW parameterization would be 5: the three prefixed exponents and the characteristic density and radius $\rho_{0}$ and $r_{0}$. Note that the 1 that is adding in the denominator is not a ``hidden parameter''. If we replaced the 1 by 2, it could be reabsorbed by redefining the free parameters as $r_{0}\rightarrow 2 r_{0}$ and $\rho_{0}\rightarrow  \rho_{0}/8$, yielding the starting point with the 1 instead of the 2.

This number of parameters is used for the computation of each $\chi^2$ per degree of freedom, and will be listed for each dark matter parametrization below in section~\ref{sec:modelprofiles}.
The impact of this choice is quite small: we will find that, given the large number of experimental points, models with more parameters often yield better fits in spite of our penalizing the number of degrees of freedom, because they can more flexibly adapt to the data.

\subsection{Use of observational data}

To study the galactic rotation curves we use the SPARC database \cite{sparc}. It contains, for a set of 175 galaxies,  rotation curves measured from the Doppler effect in the HI and H$\alpha$ lines. Furthermore, in the SPARC database the individual contributions to the rotational speed from the visible bulge $V_{\rm bulk}$, visible disk $V_{\rm disk}$ and visible gas $V_{\rm gas}$ are all estimated, based on surface photometry of galaxies at $3.6$ $\mu$m.
This allows to calculate the expected rotation curve due to the total visible matter $V_{\rm vis}\left(r\right)$ using
\begin{eqnarray}
\label{eq:vvis}
	V_{\rm vis}= \ \ \ \ \ \ \ \ \ \ \ \ \ \   \\ \nonumber \sqrt{\left|V_{\rm gas}\right|V_{\rm gas}+\Upsilon_{\rm disk} \left|V_{\rm disk}\right|V_{\rm disk}+\Upsilon_{\rm bulk}\left|V_{\rm bulk}\right|V_{\rm bulk}}\ .
\end{eqnarray}

Therein, the criterion adopted by the SPARC collaboration~\cite{lelli}  for the mass-to-light ratios $\Upsilon$ at $3.6$ $\mu$m is $\Upsilon_{\rm bulk}=1.4\Upsilon_{\rm disk}$ and $\Upsilon_{\rm disk}=0.5\,M_{\odot}/L_{\odot}$, where $M_{\odot}/L_{\odot}$ is the mass-to-light ratio for the Sun. These values come from stellar population synthesis (SPS) models, and provide the best fit for the Tully-Fisher relation \cite{McGaugh:2014,Tully:1977fu}. We adopt their extracted visible matter distributions.

\begin{figure}[h]
\centerline{
\includegraphics[width=0.4\textwidth]{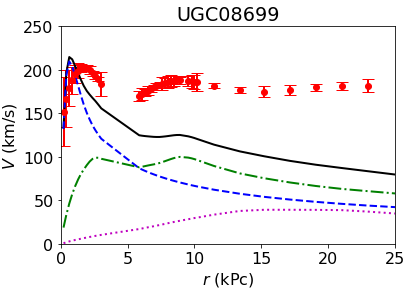}}
\caption{Rotation curve $V(r)$ of UGC08699 (circles with uncertainty bar, red online). The contributions from the bulge $V_{\rm bulk}\left(r\right)$ (dashed line, blue online), the disk $V_{\rm disk}\left(r\right)$ (dashed-dotted line, green online) and the gas $V_{\rm gas}\left(r\right)$ (doted line, pink online), as well as the sum of them $V_{\rm vis}\left(r\right)$ (solid black line with the highest value) to $V(r)$ are separately plotted. Observational data from the SPARC database \cite{sparc}. A clear mass deficit is visible and is typical of spiral galaxies.}
\label{fig:grafanalisis}
\end{figure}

A typical example rotation curve, that of UGC08699, is shown in Fig.\ref{fig:grafanalisis} together with the estimated contributions of the different matter components. 
Of note is a strong correlation between the variation of the distribution of visible matter and the variation of the $V\left(r\right)$ data. The large values of $V\left(r\right)$ at $r< 5$ kpc seem to be due to the dominant bulge. The two maxima in the disc contribution caused by the spiral arms produce an oscillation in the rotation curve at $5-10$ kpc. The gas is just important for $r>12$ kpc, where its contribution slightly counteracts the decrease of that of the bulge and disc. Thus, the variations of the rotation curves can be explained by the variations of the distribution of visible matter. However, it is insufficient to reproduce the overall level of $V\left(r\right)$ beyond $r\sim 2$ kpc, and dark matter or a modification of gravity is called for.

\subsection{Organization of the rest of the article}

In the following section~(\ref{sec:modelprofiles}) we explore the hypotheses that attempt to explain the flattening of the rotation curve, adopting 9 different models (three each for MOND variants, for a spherical dark matter halo, and for non-spherical halos) for which we calculate the expected rotation curve. To compare them, we fit the maximum possible subset from the SPARC database~\cite{sparc}, obtaining the $\chi^{2}$ per degree of freedom of each fit, and aggregating all their information in a statistical analysis.

Then in section~\ref{seccionpotencial}, fixing the density profile as function of the variable $r$ to a softened step (to gain sensitivity to the halo shape that $1/r^2$ distributions do not have),  we turn to a systematic multipole analysis of the gravitational potential and fit $V(r)$ once more. Here we find distorted and even cylindrical distributions to clearly provide better overall fits than purely spherical ones.

A variation of that same analysis is provided in section~\ref{secciondensidad}, where we incorporate a multipole expansion into the DM density function $\rho(r,\theta)$ instead of the potential (that is later calculated by numerical integration). The procedure therefore has different systematics from the earlier fits in section~\ref{seccionpotencial}; the results are compatible, though.

In section~\ref{extract_s} we turn to two independent extractions of the DM halo ellipticity from the SPARC rotation curves. As stated, we find that the quantity $\langle \log (s) \rangle$ provides a more convincing assessment, in a statistical sense, than simply $\langle s \rangle$.

Section~\ref{sec:discussion} wraps the discussion up, in particular comparing our observational data extraction to numerical simulations; and an appendix~\ref{app:oblate} collects and classifies the few galaxies among the fitted SPARC ones that contradict our statement that prolateness is the preferred explanation of rotation curves, by yielding instead an oblate fit, and in a separate table, those that seem to have observational issues or too much structure (intense oscillations) that are probably not related to dark matter.

\section{Classification of various models by $\chi^2$}
\label{sec:modelprofiles}

In this section we then proceed to contrast traditional approaches to Dark Matter at galactic scales and deformed haloes against the database. Let us first describe each of the models individually.

\subsection{ Modified Newtonian Dynamics }

MOND \cite{Milgrom:a,1983ApJ...270..371M,1983ApJ...270..384M} hypothesises the failure of Newtonian dynamics at low accelerations, of the order of a new universal constant $a_{0}$. In this regime, MONDian acceleration would be $a=\sqrt{a_{N}a_{0}}$, where $a_{N}$ is the Newtonian one. 
The flattening of rotation curves  then becomes a consequence of this transition, designed {\it ad-hoc} to avoid the need for a dark matter halo. Supposing a compact spherical visible mass distribution of mass, the outside gravitational field would change from $a_{N}=g\propto r^{-2}$ to $a=V^{2}/r=\sqrt{a_{N}a_{0}}\propto r^{-1}$, and lowering that power makes $V\left(r\right)$  constant.

To soften the nonanalyticity caused by that prescription, interpolating functions are used,
\begin{equation}\label{eq:MOND}
	\mathbf{F}=m \mathbf{a}~\mu\left(\frac{a}{a_{0}}\right)
\end{equation}
Where $\mu\left(x\right)$ is an appropriate function behaving as $\mu\left(x\right)\approx x$ at low-$x$  yielding $a=\sqrt{a_{N}a_{0}}$, but as $\mu\left(x\right)\approx 1$ at high-$x$, eliminating the correction $a=a_{N}$. Often used such functions are the so-called ``standard'' and ``simple'' proposed by Milgrom \cite{1983ApJ...270..371M} and Famaey \cite{Famaey:2005} respectively. These are given by the respective expressions $\mu_{\rm Standard}\left(x\right)=\frac{x}{\sqrt{1+x^{2}}}$ and
$\mu_{\rm Simple}\left(x\right)=\frac{x}{1+x}$
yielding 
\begin{eqnarray}\label{eq:aStand}
	a_{\rm Standard}=\frac{1}{2}\left(a^{2}_{N} \pm \sqrt{a^{4}_{N}+4a^{2}_{N}a^{2}_{0}}\right)^{\frac{1}{2}}
\\
\label{eq:aSimp}
	a_{\rm Simple}=\frac{1}{2}\left(a_{N} \pm \sqrt{a^{2}_{N}+4a_{N}a_{0}}\right) \ .
\end{eqnarray}

\begin{figure}[h]
\centering
\includegraphics[width=0.42\textwidth]{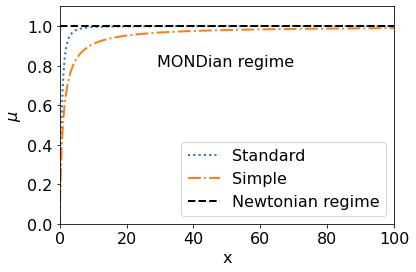}
\caption{Function $\mu\left(x\right)$ for ``simple'' and ``standard'' expressions in Eq.(\ref{eq:aStand}) and Eq.(\ref{eq:aSimp}).}
\label{fig:mongraf}
\end{figure}

As Fig.\ref{fig:mongraf} shows, the transition between the Newtonian and MOND (low $x$) regimes is sharper in the ``standard'' case, while it spreads to larger $x$ values in the ``simple'' one.  

\begin{figure}[h]
\centering
\includegraphics[width=0.42\textwidth]{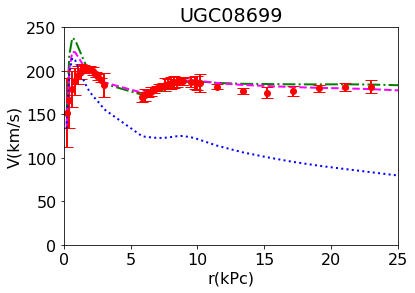}
\caption{We fit the rotation curve (circles with uncertainty bars) of UGC08699~\cite{sparc} using MOND Standard (dashed-doted, green online) and Simple (dashed, pink online) from the estimated rotation curve due to visible  matter (doted, blue online, corresponding to Newtonian mechanics). .}
\label{fig:2grafmond}
\end{figure}

The difference between the MONDian and Newtonian frameworks are noticeable in Fig.\ref{fig:2grafmond}, where we fit to UGC08699 data. Visible matter, from Eq.(\ref{eq:vvis}), with Kepler's law, yields the dotted curve that is in gross disagreement for $r>3$ kpc.
Employing instead MOND Standard, Eq.(\ref{eq:aStand}), and Simple, Eq.(\ref{eq:aSimp}), yields acceptable fits, best for $a_{0}=\left(2.24\pm 0.05\right)\cdot 10^{-10}$ m/s$^{2}$ with $\chi^{2}/N_{F}=1.8$ for MOND Standard, and $a_{0}=\left(1.63 \pm 0.04 \right)\cdot 10^{-10}$ m/s$^{2}$  with $\chi^{2}/N_{\rm F}=0.97$ for MOND Simple. 

MOND in the relevant parameter range can only with difficulty be challenged by solar-system physics. Earth's gravitational field 
would have decreased to 
the magnitude of $a_0$ at a distance  $d\sim 13.4$ a.u., so the severe consequences of the transition from Newtonian to MONDian acceleration could be seen around Pluto's orbit, $d\sim 39.48$ au; but there,  Earth's influence respect to the Sun and other bodies is negligible {\it e.g.},  $\left(a^{\rm MOND}_{\oplus}-a^{\rm }_{\oplus}\right)/a^{\rm }_{\odot}\sim 10^{-5}$, five orders of magnitude smaller than the Sun's gravitational field. For the Sun, the MOND's acceleration correction only reaches 10$\%$  ($a^{\rm MOND}_{\odot}=1.1 a_{\odot}$), at $d\approx 2000$ au, close to the hypothetical Oort cloud.

The main problem that MOND faces is that the phenomenology of the cosmos at larger than galactic scales is marginally reproduced at best: it has difficulties in reproducing cosmic microwave background anisotropies \cite{2006PhRvL..97w1301D}, the velocity dispersion and temperature profiles from galaxy clusters \cite{2001ApJ...561..550A}, and some events such as the Bullet Cluster \cite{Clowe:2004} without introducing dark matter, thus losing its main attraction.
This same observation applies to multiple attempts at modifying gravity, {\it e.g.} via the Lanczos tensor \cite{Vishwakarma:2021qay}.

\subsection{Dark matter haloes}

The rest  of models that we will fit are consistent with the most widely held hypothesis that the flattening of $V\left(r\right)$ is due to a dark matter (DM) halo, invisible at basically any wavelength.
Depending on the type of DM and its self-interactions, the formation of structure can be faster or slower.
Some simulations with light DM particles, such as ``fuzzy'' DM, produce halos that are filamentary and diffuse, and as the DM-particle mass increases, the halo shapes become lumpy \cite{Brandbyge:2017,Mocz:2019}. Hence, knowing the typical shape of halos from galaxy data can be important to address the mass (or equivalently, the Avogadro number, and speed) of DM particles.

From the SPARC velocity curves and their estimates for the visible matter component, we can derive a dark contribution to the rotation curve $V_{\rm DM}$ that we in the following use as pseudodata, 
\begin{equation}
    \label{eq:vdm}
    V^{2}_{\rm Tot}=V^{2}_{\rm DM} + V^{2}_{\rm vis} \ ,
\end{equation}

whereas when $V^{2}_{\rm vis}>V^{2}_{\rm Tot}$ we will set $V_{\rm DM}=0$.

\subsubsection{Spherical dark matter haloes ($r\equiv\sqrt{x^{2} + y^{2} + z^{2}}$)}

These have been usually assumed by analogy with other astrophysical bodies, and have the gravitational field of Eq.(\ref{eq:gravgauss}), with a mass internal to the spherical surface at $r$ of $m=4\pi\int_{0}^{r}r'^{2}dr'\rho\left(r'\right)$, yielding
\begin{equation}
\label{eq:gsph}
	g\left(r\right)=\frac{4\pi G}{r^{2}}\int_{0}^{r}{{r}'}^{2}d{r}'\rho\left({r}'\right),
\end{equation}
The rotation curve  $V\left(r\right)$ is then
\begin{equation}\label{eq:GaussS}
	V\left(r\right)=\sqrt{ 4\pi G\int_{0}^{r}{{r}'}^{2}d{r}' \frac{\rho\left({r}'\right)}{r}}
\end{equation}
To proceed, the radial profile $\rho(r)$ needs to be specified. We will employ some of the most widely used ones in the literature.

\paragraph{Navarro-Frenk-White parameterization.}

Navarro et al. \cite{Navarro:1995} carried out N-body simulations and found that their simulated halos follow an approximate density profile given by
\begin{equation} \label{eq:NFW}
	\rho\left(r\right)=\frac{\rho_{0}}{\frac{r}{r_0}\left(1+\frac{r}{r_0}\right)^{2}}\ .
\end{equation}
This parametrization has five degrees of freedom: two free parameters $\rho_{0}$ and $r_{0}$, and three ``hidden parameters'', which are the power laws of the denominator. 
Because $\rho(r)$ diverges at $r\rightarrow 0$, exhibiting a ``cusp'' nucleus, it contradicts observational data, which shows an almost constant nucleus or ``core''. This discrepancy is the so-called core-cusp problem \cite{de:Blok:2010}.  Several mechanisms such as supernovae feedback or baryonic clumps and dynamical friction have been proposed to solve the problem \cite{Del:Popolo:2016}, though the topic is still open.

\begin{figure}[h]
\centering
\includegraphics[width=0.4\textwidth]{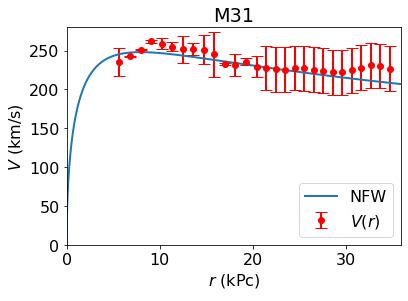}
\caption{We fit the rotation curve of M31 using the $V\left(r\right)$ from the NFW parameterization of Eq.~(\ref{eq:vNFW}). Observational data from Carignan et al. \cite{Carignan:2006}.}
\label{fig:M31}
\end{figure}

Additionally, because $\rho(r)$ decreases slowly, as $r^{-3}$ at large radii, the total mass
\begin{equation}
\label{eq:mNFW}
\begin{split}
	M & =4\pi \int_{0}^{\infty} r'^{2}dr' \frac{\rho_{0}}{\frac{r'}{r_0}\left(1+\frac{r'}{r_0}\right)^{2}} \\ 
	& =4\pi \rho_{0} r_{0}^{3} \left[ \frac{1}{1+\frac{r'}{r_{0}}}+ \log{\left(1+\frac{r'}{r_{0}}\right)} \right]_{r'= 0}^{r'\rightarrow \infty} \rightarrow \infty
\end{split}
\end{equation}
has a log divergence for $r\rightarrow \infty$. This is usually solved by introducing a cut-off in the density profile $\rho\left(r\right)\rightarrow \rho\left(r\right) \Theta\left(R_{\rm cut}-r\right)$ with a step function or a softening thereof, imposed outside the visible disk to avoid distorting the rotation curve while ensuring a finite  total mass $M$. At distances where $R_{\rm cut}$ could be noticeable, the interaction with other galaxies becomes important and asking about the mass of the individual halo stops being meaningful~\footnote{To illustrate the point, consider the halo of M31. We can guess the M31 DM fraction from cosmological values $\Omega_{M}=\Omega_{DM}+\Omega_b \approx 0.31 $  and $ \Omega_{DM}\approx 0.26$~\cite{Abbott:2019} to be around $\Omega_{DM}/\Omega_{M}\sim 0.8$. Thus,  from $M_{\rm M31}=1.5 \times 10^{12} M_{\odot}$  and with $M_{\rm DM}\sim 0.8 M_{\rm M31}$, we can estimate  $R_{\rm cut}$ from cutting off Eq.~(\ref{eq:mNFW}). The distribution parameters  are  fit to the M31 rotation curve $V\left(r\right)$ using Eq.~(\ref{eq:vNFW}),  becoming $\rho_{0}=\left(2.6 \pm 0.2 \right)\cdot 10^{-20}$ kg/m$^{3}$, $r_{0}=3.7 \pm 0.1$ kpc with $\chi^{2}/N_{\rm F}=5.2$ (see Fig.\ref{fig:M31}). Thus, the estimated cut-off radius is $R_{\rm cut} \approx 373\, r_{0}= 1373$ kpc; as this is larger than the distance between the Milky Way and M31 standing at $\approx 765$ kpc, the cut-off is beyond the validity of the concept of an isolated spiral galaxy halo.}. 
The same divergence appears in other parameterization such as the Pseudo-isothermal one (in~\ref{par:pseudoisoT} below), for example.

The rotational speed of this NFW profile can be straightforwardly calculated using Eq.~(\ref{eq:GaussS}), 
\begin{equation}\label{eq:vNFW}
	V\left(r\right)=\sqrt{\frac{4\pi G \rho_{0} r^{3}_{0}}{r}\left[\log{\left(1+\frac{r}{r_0}\right)}-\frac{\frac{r}{r_{0}}}{1+\frac{r}{r_0}}\right]}
\end{equation}
and is shown, for the case of M31, in figure~\ref{fig:M31}.

\paragraph{Pseudo-Isothermal parameterization.}
\label{par:pseudoisoT}
    
This approach requires that the DM self-interactions are sufficiently strong for the halo to thermalise, 
reaching a homogeneous equilibrium temperature. This can happen for heavy enough DM only through gravitational interaction, and may require additional, weak interactions for WIMPs~\footnote{The lack of direct DM detection sets strong bounds to possible interactions \cite{Cooley:2014}. WIMP cross sections on the nucleon are by now lower than $10^{-43}$ cm$^{2}$, making it a poor relaxation mechanism. If we limit ourselves to a purely gravitational interaction, structure formation delimits the mass somewhere between $1$ keV and $100$ GeV, where the DM halos become cuspy \cite{deVega:2012}. In the case of Strongly Interacting Massive Particles (SIMP), the Earth and Uranus heat flows \cite{Mitra:2004}\cite{Mack:2007} exclude masses from $150$ MeV to $10^{4}$ GeV, and set an upper limit on the cross section for the self-annihilation for masses from $1-10^{10}$ GeV. However, the constraints still allow a wide range of masses and interactions providing thermal equilibrium.}. This results in an isothermal sphere \cite{Weinberg:2008}, whose density profile is 
\begin{equation}
\label{eq:PIfree}
	\rho= \frac{\rho_{0}}{\left(r/r_{0}\right)^{2}}
\end{equation}
Its rotation curve $V\left(r\right)$  is then constant for any radius $V\left(r\right)=\rm const$, as can be seen in Eq.(\ref{eq:GaussS}) for Eq.(\ref{eq:PIfree}). However, because the observed rotation curves are not flat but increase at small $r$, the denominator in Eq.(\ref{eq:PIfree}) is empirically modified, without altering the wanted behaviour at large radii. Thus, 
\begin{equation}
	\rho\left(r\right)=\frac{\rho_{0}}{\left[1+\left(\frac{r}{r_0}\right)^{\alpha}\right]^{2/\alpha}}
\end{equation}
The most commonly used form of this profile incorporates $\alpha=2$, which is the so-called Pseudo-isothermal parametrization \cite{Bahcall:1980}
    \begin{equation} \label{eq:PI}
	\rho\left(r\right)=\frac{\rho_{0}}{1+\left(\frac{r}{r_0}\right)^{2}}
\end{equation}
Upon integrating Eq.(\ref{eq:GaussS}) with the density of Eq.(\ref{eq:PI}), the rotational speed obtained is 
    \begin{equation}\label{eq:vPI}
	V\left(r\right)=\sqrt{4\pi G\rho_{0} r^{2}_{0}\left[1-\frac{r_{0}}{r}\arctan{\left(\frac{r}{r_{0}}\right)}\right]}\ .
\end{equation}

\begin{figure}[h]
\centering
\includegraphics[width=0.4\textwidth]{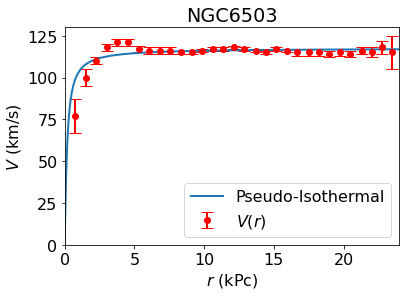}
\caption{We fit the rotation curve of NGC6503 using $V\left(r\right)$ with the Pseudo-Isothermal parameterization of Eq.(\ref{eq:vPI}). Data from SPARC~\cite{sparc}.}
\label{fig:PIgraf}
\end{figure}

This mass distribution does not suffer from the core-cusp problem, though it then desagrees with typical N-body simulations. $\rho\left(r\right)$ is almost constant at low radii, showing a core nucleus, and the second term from $V\left(r\right)$ vanishes for large r, $V\left(r\right)$ becoming contant. Hence, an isothermal distribution at large radii explains the flattened rotation curves. This can be seen in Fig.\ref{fig:PIgraf}, where the rotation curve of NGC6503 has been fitted with Eq.(\ref{eq:vPI}). 

This parameterization has three degrees of freedom: the two free parameters $\rho_{0}$ and $r_{0}$, and the ``hidden parameter'' $\alpha=2$. 

\paragraph{Einasto profile}
Einasto et al. \cite{Einasto:1989} proposed a density profile inspired by Sersic's Law \cite{Sersic:1963}, consisting on an exponential of a power law. It is usually written as 
\begin{equation} \label{eq:Ein}
	\rho \left( r \right) = \rho_{0} e^{-\frac{2}{N}\left[ \left(\frac{r}{r_{0}}\right)^{N} -1\right]}
\end{equation}

\begin{figure}[h]
\centering
\includegraphics[width=0.4\textwidth]{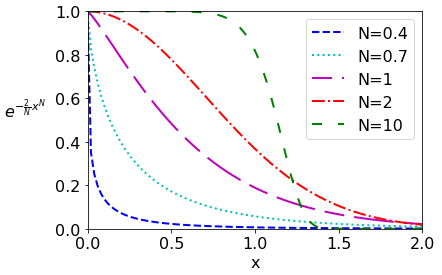}\\ 
\includegraphics[width=0.4\textwidth]{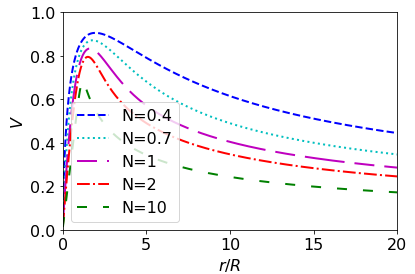}
\caption{Top: Einasto profiles, normalized to 1 at their maximum, for different values of the parameter $N$ of Eq.~(\ref{eq:Ein}). Bottom: corresponding rotation curves, with velocity normalized using $4\pi G \rho_{0} r_{0}^{2}=1$.}
\label{fig:dens_ein}
\end{figure}

The physical meaning of the  parameter $N$ can be understood from Fig.~\ref{fig:dens_ein}, where we plot the Einasto profile normalized to one at its maximum, versus $x=r/r_{0}$: $N$ controls the slope of the mass distribution. For values $N<1$, the mass lies almost entirely within the characteristic radius $r_{0}$. This is the typical mass distribution for visible matter. In fact, for $N=1/4$ we recover the de Vaucouleurs' Law~\cite{deVaucouleurs:1953} which describes the surface brightness of elliptical galaxies and bulges. For $N>1$, the mass fraction outside of $r_{0}$ increases. This mass fraction would play the role of DM and, therefore, the larger $N$ adopted, the larger contribution to the gravitational field the DM provides.

The degrees of freedom of this density profile are three: the dimensional parameters $\rho_{0}$ and $r_{0}$, and the power-law index $N$.

We know no analytical expression for the rotational speed and calculate it numerically. Typical rotation curves for some values of $N$ can be seen in the bottom plot of Fig.~\ref{fig:dens_ein}. 
They all increase until they reach a peak, from which they decay. For larger $N$, this falling becomes more abrupt, but outside it, the slope of the curve is lower. Note that there is no exact flattening for any value of $N$.

\subsubsection{Cylindrical dark matter haloes ($r\equiv\sqrt{x^{2} + y^{2}}$)}

It is not far-fetched to consider asymmetric haloes, particularly in view that visible matter is actually not symmetrically distributed, but in spiral galaxies concentrated on a disk instead. It is therefore useful to explore non-spherical halos if this would provide advantage in explaining the data.

Further, galaxy surveys, such as 6dF \cite{Jones:2009} and SDSS \cite{Alam:2015}, find a large-scale anisotropic structure or ``Cosmic Web'', with walls, filaments and voids. This structure has been reproduced in cosmological simulations \cite{Mocz:2019}\cite{2005Natur.435..629S}, where galaxies are connected by DM filaments that become clumpier as the mass of the DM particle or the intensity of their interaction increases. This suggests that the DM halo may have a filamentary contribution \cite{Llanes-Estrada:2021hnt}.  Therefore, their DM halo may be a mixture of a cylindrical contribution inherited from the filament, and a spherical contribution from the DM clump seeding the galaxy formation. 

We then explore the hypothesis that DM is distributed in elongated structures down to galactic scales: in this section, we consider three models with an exactly cylindrical halo, later we will produce interpolating parametrizations between spherical and cylindrical geometries.
Instead of the spherical radial variable $r=\sqrt{x^{2}+y^{2}+z^{2}}$, for these models we adopt cylindrical coordinates and the radial distance is, instead, $r=\sqrt{x^{2}+y^{2}}=r_{\perp }$. Both take the same value on the observable galactic plane where velocities are measured, of course, chosen as $z=0$.

The gravitational field of a cylindrical source of infinite length is easily derived through Gauss's Law, Eq.(\ref{eq:gausslaw}), by choosing a cylindrical surface of equation $x^{2}+y^{2}=r^{2}$ and $S=2\pi r L$, on which the gravitational field $g\left(r\right)$ is constant, as the contour for the integral,
\begin{equation}\label{eq:gravC}
	g\left(r\right)=-\frac{4\pi G m\left(r\right)}{2\pi r L}.
\end{equation}
Therefore, the rotation curve $V\left(r\right)$ becomes
\begin{equation}\label{eq:GaussC}
	V\left(r\right)=\sqrt{4\pi G\int_{0}^{r}r'dr' \rho\left(r'\right)}.
\end{equation}

\paragraph{Finite-width cylinder parameterization.}

    If the DM is distributed in a finite-width cylinder of constant density like Eq.(\ref{eq:FWC}), the rotation curve is still constant outside the DM halo, but inside it, the rotation curve increases linearly as in Eq.(\ref{eq:vFWC}); since the mass depends on the radius as $\frac{M}{\pi R^{2}L}=\frac{m\left(r\right)}{\pi r^{2}L}$ we have $ m\left(r\right)=M \frac{r^{2}}{R^{2}} $.
    
    \begin{equation}\label{eq:FWC}
  \rho\left(r\right) =
    \begin{cases}
      \frac{M}{\pi R^{2}L} & \text{If r $<$ R}\\
      0 & \text{If r $>$ R}
    \end{cases}       
\end{equation}
    
    \begin{equation}\label{eq:vFWC}
  V\left(r\right) = 
    \begin{cases}
      \sqrt{2 G \lambda}\,\frac{r}{R} & \text{If r $<$ R}\\
      \sqrt{2 G \lambda} & \text{If r $>$ R}
    \end{cases}       
\end{equation}

\begin{figure}[h!]
\centerline{\includegraphics[width=0.4\textwidth]{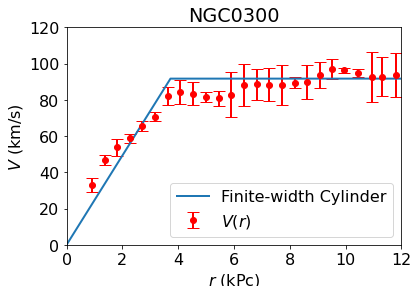}}
\caption{We fit the rotation curve of NGC0300 using the $V\left(r\right)$ of a DM finite-width cylinder of constant density, Eq.(\ref{eq:vFWC}). SPARC data from~ \cite{sparc}.}
\label{fig:cilfig}
\end{figure}

The resulting $V\left(r\right)$ perfectly captures the essence of typical measured rotation curves, as shown in Fig.\ref{fig:cilfig}.

There are two degrees of freedom for the model fit: the linear linear density $\lambda$ and the cylinder radius $R$.

\paragraph{Woods-Saxon cylinder parameterization.}
    
A drawback of the finite-width cylinder parametrization is the density discontinuity at $r=R$, Eq.(\ref{eq:FWC}), and it is interesting to introduce a ``skin'' function that provides a smoother transition. A good choice for the ``skin'' function is the one in the Wood-Saxon potential, used to describe soft nuclear edges in nuclear physics~\cite{WoodSaxon1954},
    \begin{equation}\label{eq:WS}
	\rho\left(r\right)=\frac{\rho_{c}}{1+e^{\frac{r-R}{a}}}\ .
\end{equation}

This parameterisation has three degrees of freedom: the characteristics mass density and radius of the cylinder $\rho_{c}$ and $R$, and the ``skin'' parameter $a$. 

The value of $a$ rules the smoothness of the transition at $r=R$. For $a\rightarrow 0$ the cylinder has no ``skin'' and we recover the density profile of the finite-width cylinder, while for $a\rightarrow \infty$ the ``skin'' of the cylinder is infinite and the $\rho$ becomes $r$-independent.   

\begin{figure}[h!]
\centerline{\includegraphics[width=0.4\textwidth]{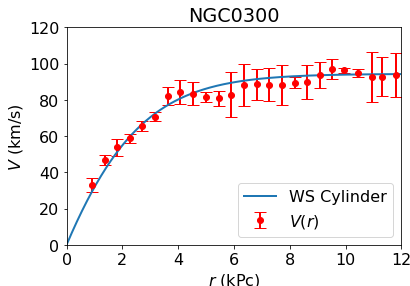}}
\caption{We fit the rotation curve of NGC0300 using $V\left(r\right)$ for the Woods-Saxon cylinder parameterization, Eq.(\ref{eq:WS}). SPARC data from~\cite{sparc}.}
\label{fig:wsfig}
\end{figure}

For this mass distribution the rotation curve $V\left(r\right)$ in Eq.(\ref{eq:GaussC}) has an analytical expression as a function of polylogarithms; however, it is  cumbersome and we rather calculate it numerically. As with the finite-width cylinder parameterization, the rotation curve $V\left(r\right)$ for finite $a$ becomes asymptotically constant as $r\rightarrow\infty$. This can be seen in Fig.\ref{fig:wsfig}, where we fit the rotation curve of NGC0300. The fit is much better than that of the finite-width cylinder, 
improving its $\chi^{2}/N_{\rm F}=3.68$  to $\chi^{2}/N_{F}=0.852$ 
 for the Woods-Saxon cylinder.

\paragraph{Generalized logarithmic potential}
 
\label{ssec:cyl_haloes_log}    

The gravitational potential outside a dense filament is a simple logarithm. A version thereof with more parameter freedom to improve data fitting is treated in this subsection: we call this version Generalized logarithmic potential because it is a generalization of James Binney's~\cite{1987gady.book.....B}, who used it to describe the gravitational field of flattened bodies (ironically,  it is a natural potential for elongated ones),
    \begin{equation}\label{eq:LogPotbad}
	\Phi\left(r\right)=\Phi_{0} \log{\left[C+ \left(\frac{r}{R}\right)^{2\alpha}\right]}
\end{equation}
except that we allow a variable $\alpha$ instead of fixing it to 1. 

This Generalized log potential is fully determined by the characteristic radius $R$, the constant $\Phi_{0}$, the power law $\alpha$, and the parameter $C$. The latter establishes the value $\Phi(0)$, and can be understood as a gauge freedom without physical impact. In fact, we can rewrite the potential as $\Phi\left(r\right)=\Phi_{0} \log{\left[1+\frac{1}{C}\left(\frac{r}{R}\right)^{2\alpha}\right]}+\Phi_{0}\log{C}$, and redefine the characteristic radius $R\,'^{2\alpha}=R^{2\alpha}/C$ yielding 
\begin{equation}\label{eq:LogPot}
	\Phi\left(r\right)=\Phi_{0} \log{\left[1+\left(\frac{r}{R\,'}\right)^{2\alpha}\right]}+\Phi_{0}\log{C}\ .
\end{equation}
    
Since the gravitational field and the mass density are derived through $\vec{g}=-\vec{\nabla}\Phi$ and $\nabla^{2}\Phi=4\pi G\rho$, after defining the new characteristic radius $R'$, both the gravitational field and mass density are independent of C. Thus, we can take $C=1$ without loss of generality (Note that if $C=0$ the predicted rotation curve is $V\left(r\right)=\sqrt{2 \alpha \Phi_{0}}$ for any radius r, since $\log\left(r\right)$ is the potential due to the straight filament of Eq.(\ref{eq:vFWC}); the finite value of $C$ makes the potential flexible enough to describe the growth of $V$ at low $r$). 
Thus, this potential is characterised by three free parameters:  $R$,  $\Phi_{0}$ and the power law $\alpha$. 
    
Poisson's equation provides the mass density profile
\begin{equation}\label{eq:maslogpot}
	\rho\left(r\right)= \frac{\alpha^{2} \Phi_{0}}{\pi G}\frac{1}{r^{2}}\frac{r^{2\alpha}/R^{2\alpha}}{\left(1+r^{2\alpha}/R^{2\alpha}\right)^{2}}
\end{equation}
that imposes some parameter restrictions due to the $\rho>0$ positiveness condition, namely $\alpha>0$ and $\Phi_{0}>0$. 

\begin{figure}[h]
\centering
\includegraphics[width=0.5\textwidth]{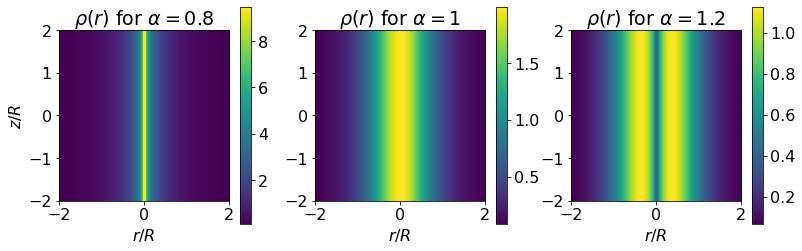}
\caption{Density plots of $\rho\left(r\right)$ in Eq.(\ref{eq:maslogpot}) for $\Phi_{0}=2\pi G R^{2}$ and different values of $\alpha$, showing a cusp (left), a core (middle) and a shell-like (right) cylindrical haloes. }
\label{fig:imshowlogden}
\end{figure}

Fig.\ref{fig:imshowlogden} shows $\rho\left(r\right)$ for $\Phi_{0}=2\pi G R^{2}$ and different values of $\alpha$. For $0<\alpha<1$ we find cusp halos (the plot has the appearance of a narrow slit), for $\alpha=1$ they became softer and a core appears, while for $\alpha>1$ the core disappears and the halo is shell-like. Curiously, this later case has been observed in large-scale structure formation simulations for warm, for hot and for fuzzy DM \cite{Brandbyge:2017,Mocz:2019}.

In fact, we can intuitively relate $\alpha$ to unknown microscopic properties such as the mass of the DM particles or the intensity of their interaction. An initial velocity dispersion of the DM particles could be smaller than at equilibrium, $\sigma_{V}/\sigma_{V}^{\rm eq} \ll 1 $. Their interactions would widen that dispersion: Less energetic particles populate small $r$ orbits, while  more energetic ones will be found at large radii. Therefore, $\rho(r)$ in principle conveys information about the strength of the interaction, {\it i.e.} the DM particle mass for the gravitational interaction, or its charge associated to other interactions. Heavy (or strongly interacting) DM particles concentrate at the galactic centre and generate cusp halos, while light (or weakly interacting) DM particles spread to larger radii, producing, in an extreme case, shell-like halos. This means that $0<\alpha<1$ corresponds to heavy (or strongly interacting) DM particles, $\alpha>1$ to light (or weakly interacting) DM particles, and  $\alpha \simeq 1$ provides the transition between the two regimes.

The predicted rotation curve is given by 

\begin{equation}\label{eq:vLogPot}
	V\left(r\right)=\sqrt{2\alpha \Phi_{0} \frac{r^{2\alpha}/R^{2\alpha}}{1+r^{2\alpha}/R^{2\alpha}}} \ .
\end{equation}

\begin{figure}[h]
\centering
\includegraphics[width=0.42\textwidth]{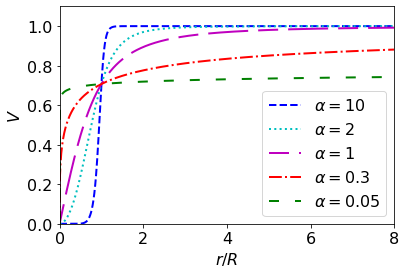}
\caption{Shape of the rotation curves of the generalized logarithmic potential for different values of $\alpha$, where the velocity has been normalised to $V\left(R\right)=1/\sqrt{2}$.}
\label{fig:vlogvalues}
\end{figure}

Note that at large radii $r/R\gg 1$,  $V\left(r\right)\rightarrow\sqrt{2\alpha \Phi_{0}}$ becomes constant for fixed $\alpha$. Normalized rotation curves for some values of $\alpha$ can be seen in Fig.\ref{fig:vlogvalues}.

\subsection{Data and Analysis}
\label{ssec:datosoli}

We now proceed to compare the nine model approaches described so far, 
ordering them on the basis of their ability to reproduce the observed rotation curves. For this purpose, we perform a least-squares fit of each rotation curve of the SPARC database \cite{sparc}, minimising $\chi^{2}$  as a function of the free parameters of the model, 
\begin{equation}
\label{eq:chi2}
	\chi^{2}=\sum_{i=1}^{N}\frac{\left[V^{\rm obs}_{i}-V^{\rm th}\left(r_{i}\right)\right]^{2}}{\Delta V^{\rm obs\ 2}_{i}}\ .
\end{equation}
Therein,  $V^{\rm obs}_{i} \pm \Delta V^{\rm obs}_{i}$ is the experimental rotation curve with its uncertainty, and $V^{\rm th}$ the model prediction.

For the six models that involve Dark Matter (three with spherical and three with cylindrical geometry) we employ for $V_{\rm obs}$ the pseudo data $V_{\rm DM}$ derived in Eq.~(\ref{eq:vdm}), so that only the dark matter contribution to the squared velocity is fit.
For the three models without DM (pure Newtonian mechanics with the visible matter and MOND) we employ the measured velocity $V_{\rm obs}=V_{\rm Tot}$, without subtracting any visible contribution (as is obviously necessary in the earlier ones with DM). 
Also, for these models without DM we compute $V_{\rm th}$ from the estimated visible $V_{\rm vis}$: $V^{2}_{\rm vis}/r$ plays the role of the Newtonian acceleration $a_{\rm N}$ in Eqs.~(\ref{eq:aSimp}) and~(\ref{eq:aStand}).

However, since each model approach has a different number of parameters, we calculate the number of degrees of freedom (d.f.) for each of the galaxies, calculate $\chi^{2}/N_{\rm F}$, with
\begin{equation}
\label{eq:Nfreedom}
	N_{\rm F}=N_{\rm points}-\rm d.f.
\end{equation}
being the difference between the number of points on a given SPARC rotation curve and the number of free plus hidden parameters of the tested model.

Because $\chi^{2}/N_{\rm F}$ needs to be positive, we need $N_{F}>0$. Because we want to test all the models against the same sample of galaxies for a fair comparison, we only use those galaxies whose observational rotation curves exceed 5 points, which is the largest number of parameters of any of the examined models (saturated by the NFW parameterization). This reduces the sample of 175 rotation curves to 164. Besides, we are forced to exclude the rotation curve of UGC01281, since the estimated $V_{\rm vis}$ becomes complex at small radii, reflecting some observational analysis issue, leaving a total of 164 rotation curves.

For each rotation curve $V\left(r\right)$ of this subset and each of the models, we compute the optimal  $\chi^{2}/N_{\rm F}$ over the model parameter space. The minimization is carried out employing the well established CERN's Minuit algorithm as implemented by standard python libraries~\cite{iminuit}, and a pass over the entire galaxy database runs in a few hours in a standard departmental Linux cluster.

Each galaxy then yields a $\chi^2/N_F$ ranking of the nine model approaches  from 1, at the smallest $\chi^{2}/N_{\rm F}$, to  the $9^{\rm th}$ having the largest such (as summarized in Table~\ref{table:summarymodels}): the lower the model ranking number, the better the overall fit. In the case where a galaxy assigns two or more models essentially the same $\chi^{2}/N_{\rm F}$ (which is calculated to four digits to minimize this possibility), the rank assigned to all those with a degenerate value is the group's average, {\it e.g.} if two of them coincide on the smallest value of $\chi^{2}/N_{\rm F}$, they both receive the rank $(1+2)/2=1.5$.

\begin{figure}[h!]
\centering
\includegraphics[width=0.7\textwidth]{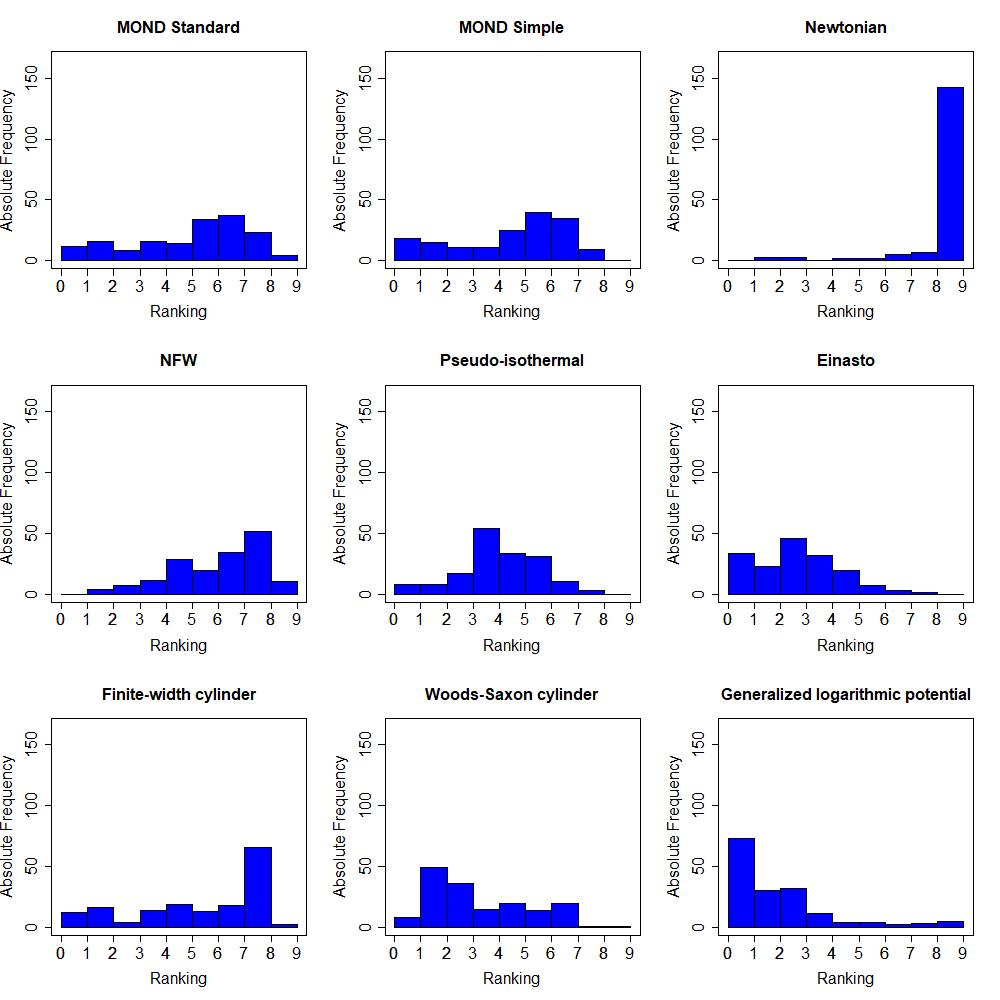}
\caption{ Histograms of the distribution of model rankings for $\chi^{2}/N_{\rm F}$ for each parameterization.}
\label{fig:Hist3x3}
\end{figure}

We show histograms of these rankings in Fig.~\ref{fig:Hist3x3}, 
where the height of the bars represents the number of galaxies that assign the given ranking (along the $OX$ axis) to the model in the given plot.
Further, in Table~\ref{table:summaryolianalysis} we provide the parameters of the ranking distributions over the galaxy population, namely their means $\bar{x}$, medians Med, standard deviations $\sigma$, and median absolute deviations MAD.

\onecolumngrid
\begin{center}
\begin{table} [h] 
\begin{center}
\begin{tabular}{ |c|c|c|c|c|c| } 
 \hline
 Model &  Mass Distribution &  $V\left(r\right)$ & d.f. & $N_{F.P.}$ & Flat. \\ \hline
 Newtonian & Visible only & $V_{\rm vis}$ & 0 & 0 & No \\ \hline
 MOND Standard & Visible only & $V^{2}_{\rm vis}/r$ in Eq.(\ref{eq:aStand}) & 2 & 1 & Yes \\ \hline
 MOND Simple & Visible only & $V^{2}_{\rm vis}/r$ in Eq.(\ref{eq:aSimp}) & 2 & 1 & Yes \\ \hline
 NFW & DM Eq.(\ref{eq:NFW}) & Eq.(\ref{eq:vNFW}) & 5 & 2 & No  \\ \hline
 Pseudo-Isothermal & DM Eq.(\ref{eq:PI}) & Eq.(\ref{eq:vPI}) & 3 & 2 & Yes \\ \hline
 Einasto & DM Eq.(\ref{eq:Ein}) & Numerically & 3 & 3 & No \\ \hline
 Finite-width cylinder & DM Eq.(\ref{eq:FWC}) & Eq.(\ref{eq:vFWC}) & 2 & 2 & Yes \\ \hline
 Woods-Saxon cylinder &  DM Eq.(\ref{eq:WS}) & Numerically & 3 & 3 & Yes \\ \hline
 Generalized logarithmic potential &  DM Eq.(\ref{eq:maslogpot}) & Eq.(\ref{eq:vLogPot}) & 3 & 3 & Yes \\ \hline
\end{tabular}
\end{center}
\caption{Summary of the model approaches examined against the galaxy database: mass distribution, rotation curve $V\left(r\right)$, number of degrees of freedom, number of free parameters, and whether the expected rotation curve $V\left(r\right)$ flattens out at large $r$.}
\label{table:summarymodels}
\end{table}
\end{center}

\begin{table} [h] 
\begin{center}
\begin{tabular}{ |c|c|c|c|c| } 
 \hline
 Ranking order & Model & $\bar{x}\pm \sigma$ & Med $\pm$ MAD  \\ \hline \hline
 1 & Generalized log potential (cylindrical) & $2.4\pm 1.9  $ & 2  $\pm$ 1 \\ \hline
 2 & Spherical Einasto & $3.1\pm  1.6 $ & 3 $\pm$  1  \\ \hline
 3 & Woods-Saxon cylinder & $3.7\pm 1.9 $ & 3 $\pm$  1  \\ \hline
 4 & Pseudo-Isothermal & $4.5\pm 1.5$ & 4 $\pm$  1 \\ \hline
 5 & MOND Simple & $4.9 \pm 2.1 $& 6  $\pm$ 1  \\ \hline
 6 & MOND Standard & $5.4\pm 2.2 $ & 6 $\pm$ 1 \\ \hline
 7/8 & Finite-width cylinder & $5.8\pm 2.4 $ & 7 $\pm$  1  \\ 
 7/8 & Spherical NFW & $6.5\pm1.7  $ & 7 $\pm$ 1  \\ \hline
 9 & Newtonian & $8.6\pm 1.4 $ & 9 $\pm$  0 \\ \hline
\end{tabular}
\end{center}
\caption{Measures of centrality and dispersion of the model ranking: we give the average with standard deviation and the median with median absolute deviation of the position in which the galaxy fits prefer each of the models. Clearly, purely Newtonian physics with visible matter only yields the worse overall fits. The (purely spherical) Einasto and (purely cylindrical) logarithmic dark matter potentials yield the best fits, of comparable quality to each other, with other spherical and cylindrical approaches, having less parameters, following in the given order.}
\label{table:summaryolianalysis}
\end{table}

\newpage

\twocolumngrid

The diversity in the measured rotation curves, not all of which extend far enough to 
flatten out as seems to be the norm, makes different model approaches better suited for different galaxy subsets. Such dispersion can be exposed by overall ``agreement'' vs. ``disagreement'' tests (useful in other contexts to understand the opinion of jurors or committee members choosing among several options). 
For example, ascertaining such agreement among different galaxies we have performed Kendall's W test~\cite{10.1214:aoms:1177732186}. 
This is based on the matrix $r_{i,j}$ containing the rank assigned to model $i$ by each rotation curve $j$.
The statistical criterion is based on the number 
\begin{equation}
    W=\frac{12 S}{\left(n^{3}-n\right)}\ ,
\end{equation}
where $n=9$ is the number of models to be compared.
The numerator $S$ is constructed from the average rank assigned to each model by each of the $m=164$ usable rotation curves with sufficient measured data points, $R_{i}=\frac{1}{m}\sum_{j=1}^{m}r_{i,j}$. 
This is then averaged over all the models $\bar{R}=\frac{1}{n}\sum_{i=1}^{n}R_{i}$ and a quadratic deviation constructed therewith as
$S=\sum_{i=1}^{n}\left(R_{i}-\bar{R}\right)^{2}$.

Kendall's $W$ values range between $0$ (for largest disagreement among galaxies) to $1$ (for full agreement: all would favor the same model). We obtain $W=0.013$, and after correcting for tied rank, $W=0.014$; because it is quite small, a large ``dispersion'' in the rankings, visible in table~\ref{table:summaryolianalysis}, is suggested. This amply  justifies the further work in this article.

In spite of this dispersion, we have some statistical confidence that we may order the models as in table~\ref{table:summaryolianalysis} from a computation of 
the $U$ test of Wilcoxon-Mann-Whitney \cite{10.1214:aoms:1177728261}.

This is a nonparametric test that compares the medians according to the following criterion: given two samples of $n_{1}$ and $n_{2}$ elements, we order the total set giving each element the rank $r_{j,i}$ (the rank in the total set of the element $i$ from the sample $j$), from which we calculate the parameter

\begin{eqnarray}
    U_{2}=\sum_{i=1}^{n_{2}}r_{i,2} -\frac{n_{2}\left(n_{2}+1\right)}{2}
\end{eqnarray}

We reject the null hypothesis that the median of sample $2$ is smaller than the median of the sample $1$, symbolically $H_{0}:~ {\rm Med}_{\rm 2}< {\rm Med}_{\rm 1}$, if $U_{2}<C_{n_{1},n_{2},p}$; where $C_{n_{1},n_{2},p}$ are tabulated coefficients and $p$ is the p-value. 

Its application to the problem at hand is as follows: if the $p$-value is $p<0.05$, so that we accept instead the alternative hypothesis of the median of the model $i$ is bigger than the median of the model $j$  $H_{1}:~ {\rm Med}_{\rm j}> {\rm Med}_{\rm i}$ for a confidence level of 95$\%$ 
(transitivity is self-evident, once model $i$ drops below model $j$, and $j$ below $k$, $i$ ranks below $k$). 

A clear conclusion is that the worst model is the purely Newtonian rotation curve based on visible matter, where the computed $p$-values with respect to the other models are all less than $p<2.2\times 10^{-16}$, rejecting the null hypothesis, and making it significantly the worst. This is very strong statistical evidence that visible matter is not sufficient to explain the measured rotation curves.

However, we are unable to rank Navarro-Frenk-White's spherical parametrization versus a finite-width uniform cylinder ($p= 0.053$), but we can say that these provide worse fits than MOND Standard ($p=3.7\times 10^{-6}$ and $p=0.0056$, respectively). In turn MOND Standard is below MOND Simple ($p=0.011$), and this one in turn concedes to the spherical Pseudo-Isothermal parametrization ($p=0.00083$). The Woods-Saxon cylinder lies above this in fit quality ($p=9.07\times 10^{-06}$), but
below Einasto's spherical one ($p=0.0065$), which finally is trumped by the Generalized Logarithmic Potential ($p=1.8\times 10^{-07}$), a cylindrical parametrization. 

With these tests completed, we obtain the classification in table \ref{table:summaryolianalysis}, and notice that spherical and cylindrical parametrizations are interspersed.

As an aside, given that the  Generalized Logarithmic Potential seems to provide the best fit, it is interesting to extract from the fitted galaxy sample
the distribution of its parameter $\alpha$ in Eq.~(\ref{eq:LogPot}) and its attending uncertainty. For this extraction we add the mild requirement (that only exclude two of the sampled galaxies, NGC4085 and UGC06787) that $\Delta \alpha$ is computable and different from $0$, leaving a subset of 162 SPARC galaxies for the fit. 

\begin{figure}[h]
\centering
\includegraphics[width=0.4\textwidth]{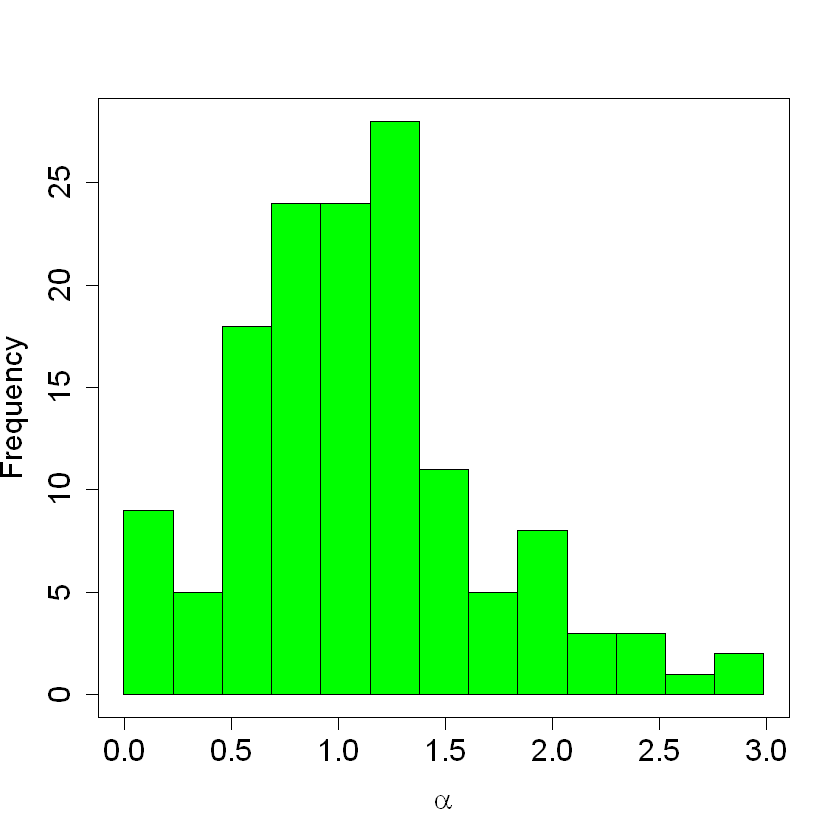}
\caption{Histogram of the distribution of the parameter $\alpha$ in Eq.~(\ref{eq:LogPot}), for the Generalized Logarithmic Potential (very similar to that outside a cylindrical source of constant density for large $r$), the one that best describes the galaxy database (table~\ref{table:summaryolianalysis}).
We have excluded a few extreme (outlier) values, not shown, from the fit. 
The distribution is quite peaked, and yields $\alpha=0.456\pm 0.003$.}
\label{fig:hist_alpha}
\end{figure}

The distribution of $\alpha$ is shown in the histogram of Fig.\ref{fig:hist_alpha}: not included there are those extreme values that can be tagged as outliers. These are characterized in terms of the $\rm i$th quartile  $\rm Q_{i}$ and the interquartile range  $\rm IQR$ by the conditions
\begin{eqnarray}
\alpha &<& \rm Q_{1} - 1.5 ~IQR=-0.78 \nonumber \\
\alpha &>&  \rm  Q_{3} + 1.5~\rm IQR=3.3\ .
\end{eqnarray}

 The resulting 162-galaxy population's ``central'' value for the variable $\bar{\alpha}$  and its uncertainty $\Delta \alpha$ is obtained by minimising the statistical estimator
\begin{equation}
\label{eq:chi2_2}
	\chi^{2}=\sum_{i=1}^{N}\frac{\left[\alpha^{\rm fit}_{i}-\alpha\right]^{2}}{\Delta \alpha^{\rm fit\ 2}_{i}}\ .
\end{equation}

The outcome is $\bar{\alpha}\pm \Delta \alpha =0.456\pm 0.003$. 

Comparing it with Fig.~\ref{fig:imshowlogden} we see that this small value of $\alpha$ implies that the mass density of the cylindrical DM halo would have  a very pronounced peak at $r\sim 0$ (a central filament-like structure, analogous to the ``cusp'' of spherical simulations), 
 suggesting that the DM particles are heavy or strongly interacting (see discussion above Fig.\ref{fig:imshowlogden} in section \ref{ssec:cyl_haloes_log}).

As seen also in figure~\ref{fig:Hist3x3}, we find that both cylindrical geometries and spherical geometries can account for the $V(r)$ rotation curve, and that the deciding factor, for relatively simple models, is the number of parameters that allow a better $\chi^2$.

To gain closer understanding we wish to examine intermediate geometries between the extreme spherical and cylindrical one with varying prolateness, and even allow the fits to eventually produce the opposite, oblate DM distributions when necessary. 

The systematic way to address this problem is to employ a multipole expansion. This we describe in the next two sections. In Sec.\ref{seccionpotencial} 
we are going to directly expand the potential shape and use the resulting coefficients, that are expressible in terms of the underlying mass-density distribution.
In Sec.~\ref{secciondensidad} we will instead expand the density and compute the potential therefrom. The two methods have different systematics, but yield similar results.

\newpage

\section{Multipole analysis of the gravitational galactic potential}
\label{seccionpotencial}

\begin{figure}
  \centering
  \includegraphics[width=.9\linewidth]{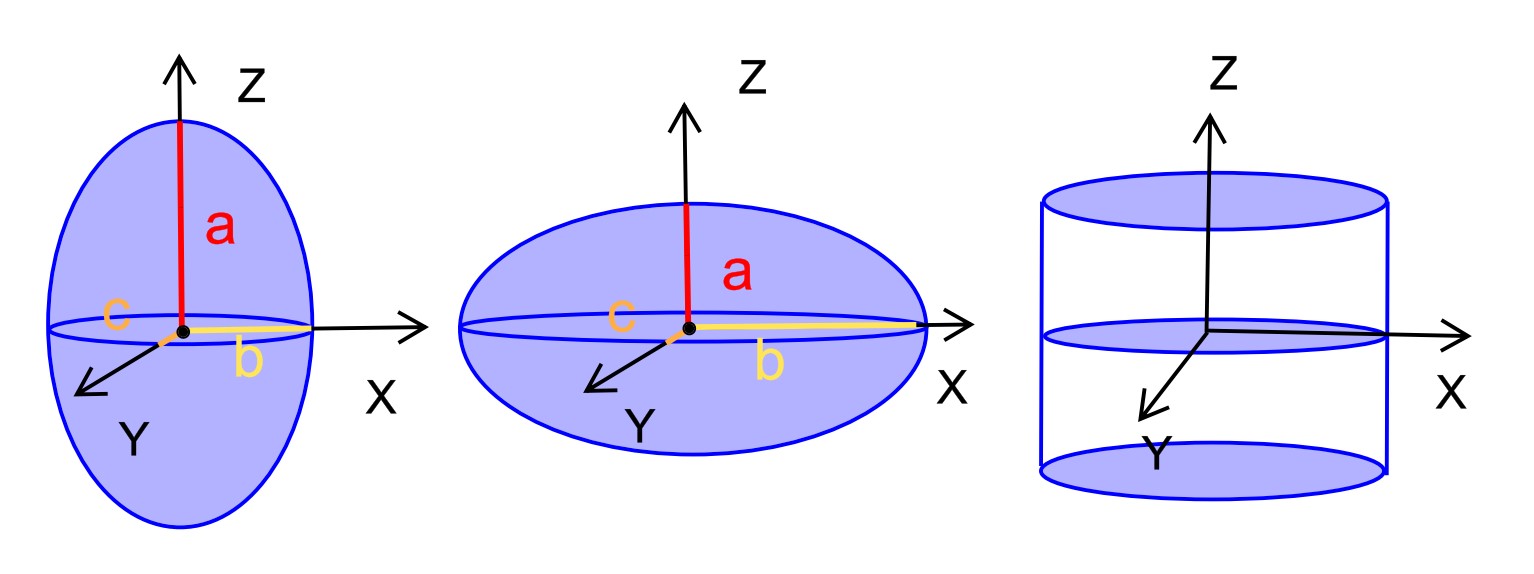}
  \caption{Variables used to describe typical elongated halo geometries. Left: interpolating prolate ellipsoidal mass distribution with $a$ the major and $b$, $c$ the minor semiaxes. Middle: extrapolating ellipsoidal oblate mass distribution with $a$ the minor and $b$, $c$ the major semiaxes. Right: cylindrical distribution as the extreme case opposed to a spherical one.
    \label{haloelongadoforma}}
\end{figure}

In this section we perform a direct multipole expansion of the potential; this is to avoid bias by choosing one particular density distribution as function of the radial-like scale. We will use it to fit the galactic rotation curves of SPARC's database assuming an ellipsoidal shape of the haloes (see Fig.~\ref{haloelongadoforma}). Then we will obtain the degree of ellipticity that these objects present, with the notation of Table~\ref{triaxiality} employed throughout.
\begin{table}[h!]
    \centering 
    \begin{tabular}{|c|c|}\hline
        \textbf{Halo}     &  \textbf{Axes relation} \\ \hline
        Prolate  &  $a>b \simeq c$\\\hline
        Oblate   &  $a \simeq b> c$\\\hline
        Triaxial &  $a>b>c$ \\ \hline
    \end{tabular}
    \caption{ Convention for the relative size of the characteristic lengths $a$, $b$, $c$ along the three principal axes of an ellipsoidal halo.  See Figure~\ref{haloelongadoforma}.}
    \label{triaxiality}
\end{table}

The gravitational potential generated at point $\bf r$ by an ellipsoidal density distribution can be obtained as the sum of the potentials of spherical shells of mass $dM({\bf r})=\int d^3 r' \rho({\bf r'})$ (see top sketch in Fig.~\ref{momentodeinerciaMEP} below), as
\begin{equation}
    \Phi(\textbf{r})=-\frac{G}{|r|}\int^{|r|}_0 d^3r' \: \rho({\bf r'})-G\int^\infty_{|r|} d^3r' \frac{ \rho({\bf r'})}{|r'|} \ .
    \label{sphericalshelling}
\end{equation}

Because of the lack of spherical symmetry of the mass distribution, the ``outer'' layers at $|r'|>|r|$ contribute. To systematically treat this deviation from sphericity,  we will deploy a multipole expansion in terms of $Y_l^m(\theta,\varphi)$ \cite{BinneyTremaine} 
demanding that the expansion coefficients encode  the residual symmetries of the ellipsoid. These are the axial symmetry around the $OZ$-axis (that fixes the second multipole index as $m=0$) and the reflection-symmetry on the galactic plane (that restricts $l$ to be even). Thus,
\begin{equation}
    \Phi(\textbf{r})=- G \sum_{l\ {\rm even}, m=0} P_{ml}(\theta,\phi)\left[\frac{I_{l0}(r)}{r^{l+1}}+r^lQ_{l\geq 2,m=0}(r)\right]
    \label{expansion}
\end{equation}
Here $I_{lm}$ and $Q_{lm}$ represent the coefficients of the internal- and external-layer contributions, respectively, obtained upon integrating the density in the respective domain over the corresponding Legendre polynomials  $P_{l0}(\theta) = \sqrt{4 \pi /(2l+1)}Y_{l0}(\theta)$,
\begin{equation}
    I_{l0} (r>r') = \int d\Omega'\int^r_0 dr' \:(r')^{l+2} \rho(\textbf{r}') \, P_l(\theta') 
    \label{interna}
\end{equation}
\begin{equation}
    Q_{l0} (r<r') = \int d\Omega'\int^\infty_r dr' \:  (r')^{1-l} \rho(\textbf{r}') \, P_l(\theta') \ . 
\end{equation}
(The upper integration limit in $Q_{l0}$ is of course a maximum $R$ when implemented on a computer; the external, spherical $Q_{00}$ does not contribute because of Gauss's theorem, so it is best left out of the sum from the start to diminish numeric noise.)

We extend the multipolar expansion to the monopolar $l=0$, quadrupolar $l=2$ and hexadecapolar $l=4$ terms.
We then obtain the rotation velocity from the resulting potential in Eq.~(\ref{expansion}) as
\begin{equation}
    V^2=-r\frac{\partial \Phi(\textbf{r})}{\partial r}\ .
    \label{velocidadd}
\end{equation}

To examine the geometry with this analysis, independent of that in 
section~\ref{sec:modelprofiles},  we need to adopt any one of the reasonable halo density profiles
as function of the distance scale.
We have chosen two of the simplest models: the first is simply a two-parameter step distribution

\begin{equation}
\rho(r) = \rho_0\Theta(R-r)
\end{equation}

with a constant inner density that suddenly drops to zero outside of $R$. 
As a second parametrization, we have chosen a Woods-Saxon density profile,
a well studied function that is used in several fields of physics to represent a
core followed by a decrease whose value decays away from the gravitational source in a parametrically controlled way (it is also functionally identical to the Fermi-Dirac distribution), 
\begin{equation}
    \rho_{W-S} (r) = \frac{\rho_0}{1+e^{(r-R)/a_0}}\ .
    \label{saxonformula}
\end{equation}

Each of these two has been employed with each of the first three orders of the expansion in Eq.~(\ref{expansion}), and then also with a purely cylindrical distribution to yield a total of eight different parametrizations listed in Table~\ref{modelostable}.

\onecolumngrid

\begin{figure}
  \centering
  \includegraphics[width=.3\linewidth]{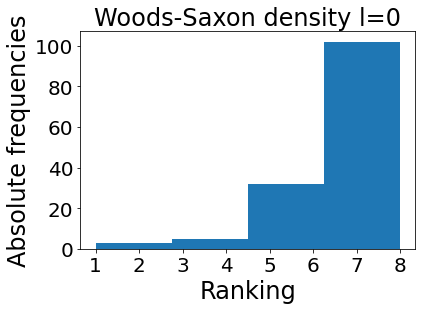}
  \includegraphics[width=.3\linewidth]{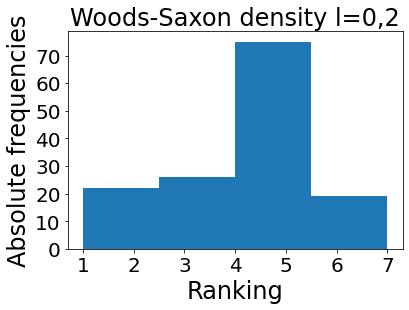}
  \includegraphics[width=.3\linewidth]{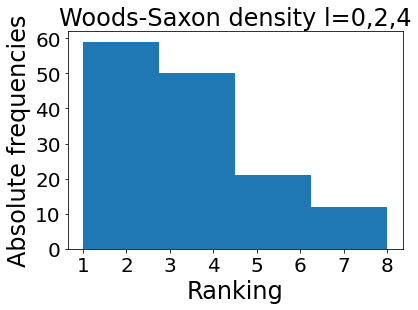}
  \includegraphics[width=.3\linewidth]{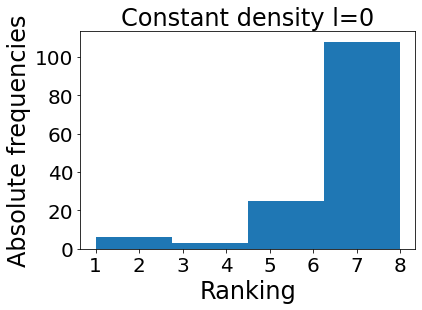}
  \includegraphics[width=.3\linewidth]{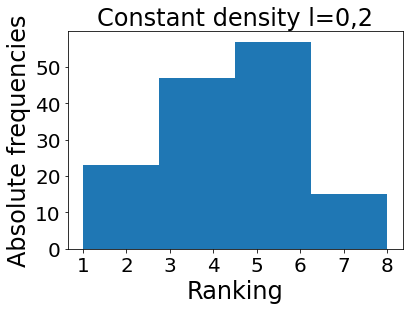}
  \includegraphics[width=.3\linewidth]{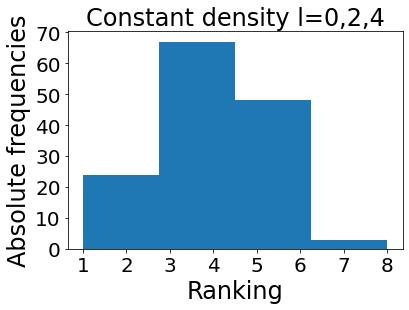}
  \includegraphics[width=.3\linewidth]{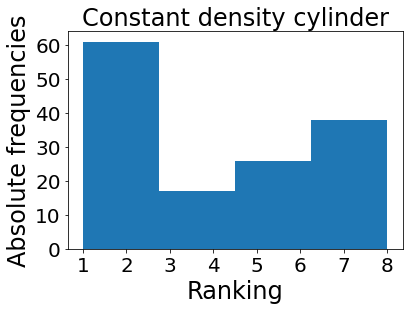}
  \includegraphics[width=.3\linewidth]{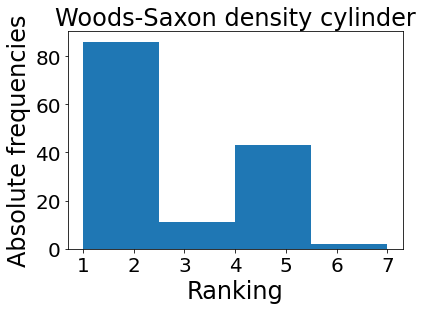}
  \caption{Histograms exposing the distribution of $\chi^{2}/N_{\rm F}$ rankings  for each parameterization in Table~\ref{modelostable} (mapping out a multipole expansion of the potential). The $OY$ axis is the number of galaxies that classify the calculation in each box in the corresponding bin. It is clear that both spherically-symmetric models ($l=0$, top and middle left) are ranked in the worst tier (6, 7 or 8) by a large majority of galaxies whose $V(r)$ is fit. On the other hand, it is not clear that we have a successful extraction of the $l=4$ multipole, so we  concentrate on $l=2$. The cylindrical fits (bottom row) are rather good too.
   }
  \label{graficosMEP}
\end{figure}

\newpage

\twocolumngrid

Each galaxy $j=1,..., 109$
(those with acceptable rotation curves that do not oscillate violently, see discarded galaxies in Table~\ref{galaxias_descartadas}, and sufficiently many measured points to be usable for this analysis)
assigns each parametrization $i=1,..,8$ a rank $R_{ij}$ based on ordering the fit $\chi^2$ from smaller to larger values. The rank can take values from 1 to 8, where 1 describes the best parametrization. We can obtain a global rank for each approach calculating the mean value $\bar{R_i} = \frac{1}{N}\sum^m_j R_{ij}$ or the median of the individual galaxy ranks. 
These mean and median are also listed in Table \ref{modelostable}.

\begin{table}[h!]
    \centering
    \begin{tabular}{|c|c|c|c|c|}\hline 
       \textbf{Angular} & \textbf{Density} & $N_{F.P.}$ & \textbf{Median} & $\bar{R_i} \pm \sigma$ \\ 
       \textbf{shape} & \textbf{profile} & & \textbf{$\pm$ MAD} & \\ \hline 
        Cylinder & Woods-Saxon     & 4 & 2.0$\pm$1.4 & $2.8 \pm 1.6$ \\ \hline 
        $l=0$, 2, 4 &  Woods-Saxon & 7 & 3.0$\pm$1.5 & $2.8 \pm 2.1$ \\ \hline
        $l=0$, 2 & Woods-Saxon     & 5 & 4.0$\pm$1.1 & $3.7 \pm 1.4$ \\ 
        $l=0$, 2, 4 &  Constant    & 4 & 4.0$\pm$1.1 & $3.9 \pm 1.4$ \\  
        $l=0$, 2 & Constant        & 3 & 4.0$\pm$1.6 & $4.4 \pm 1.9$ \\ \hline 
        Cylinder & Constant        & 3 & 5.0$\pm$2.3 & $4.7 \pm 2.5$ \\ \hline
        $l=0$ & Constant           & 2 & 7.0$\pm$0.8 & $6.9 \pm 1.5$ \\
        $l=0$ & Woods-Saxon        & 2 & 7.0$\pm$0.9 & $6.9 \pm 1.3$ \\ \hline 
        
    \end{tabular}
    \caption{Angular-dependence parametrization, density profile as function of the distance to the galactic center, number of fit parameters, median value, mean value, and standard deviation.  (This analysis is based on 109 of the 175 galaxies; only those with at least 7 measured points and with good data quality have been employed.)}
    \label{modelostable}
\end{table}
As the table shows, a purely cylindrical potential (and thus, the entailed dark matter distribution) and those with higher multipoles (distorting the spherical symmetry) seem to perform better than those nearly spherical shapes. 

The full histogram distribution from which the table is extracted is also shown in Fig.~\ref{graficosMEP}.  
It is patent to the eye that the best fitting angular distributions are those with a filamentary distribution of dark matter (cylindrical distribution or higher order terms $l=0$, 2, 4). In galaxies where the rotation curve lies flat from the beginning, such as in  Fig.~\ref{fig:rotNGC3198pot} shortly, higher order multipoles will be needed to fit the rotation curve. The worst fits are obtained with  a spherical distribution of dark matter. We expect this to be a generic feature of most DM density profiles excepting those that are near the isothermal one, $\rho \propto 1/r^2$.

\begin{figure}[h!]
    \centerline{\includegraphics[width=0.4\textwidth]{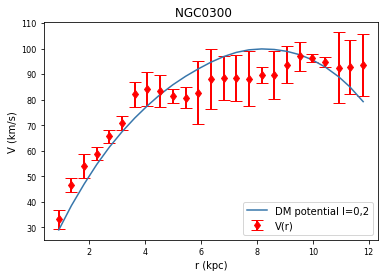}}
    \centerline{\includegraphics[width=0.4\textwidth]{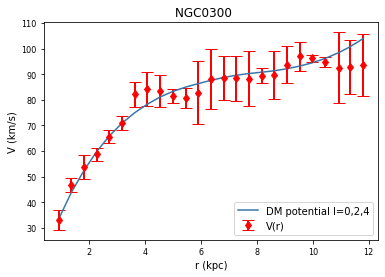}}
    \caption{We fit the rotation curve of NGC0300 using the $V\left(r\right)$ of the multipole expansion of DM density distribution, Eq.(\ref{expansion}). SPARC data from~ \cite{sparc}.}
    \label{fig:rotNGC0300pot}
\end{figure}
\begin{figure}[h!]
    \centerline{\includegraphics[width=0.4\textwidth]{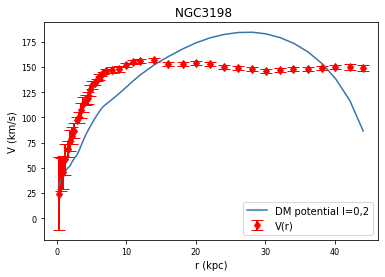}}
    \centerline{\includegraphics[width=0.4\textwidth]{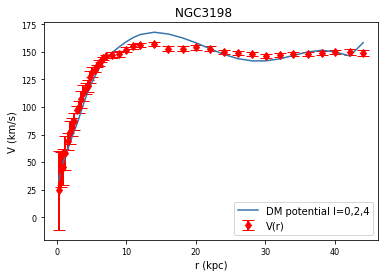}}
    \caption{We fit the rotation curve of NGC3198 using the $V\left(r\right)$ of the multipole expansion of DM density distribution, Eq.(\ref{expansion}). SPARC data from~ \cite{sparc}. We realise that for a correct fit we need higher order multipoles $l=$0,2,4...}
    \label{fig:rotNGC3198pot}
\end{figure}

\subsection{Moments of inertia and length of the semiaxes}

A traditional and physically transparent way of expressing the potential of a nonspherical body is in terms of the integrated moments of the mass density.
The internal contribution to  $\Phi$ by the quadrupolar term in Eq.~(\ref{interna}),
\begin{eqnarray}
    I_{20} &=& \int d^3r' (r')^2 \rho(\textbf{r}')P_2(\theta) \nonumber \\ &=&
    \sqrt{\frac{4 \pi}{20}}\int d^3r' (r')^2 \rho(\textbf{r}')\:(3\cos^2\theta-1)\ .
    \label{integrando}
\end{eqnarray}
The moments of order 2 are best expressed in terms of the moments of inertia, as usual when dealing with  deformations of celestial bodies~\cite{goldstein,jupyter}, 
\begin{eqnarray}
    J_1 &:=& \int dV \rho \cdot (y^2+z^2) \nonumber \\
    J_2 &:=& \int dV \rho \cdot (x^2+z^2) \nonumber \\
    J_3 &:=& \int dV \rho \cdot (y^2+x^2)\ .
    \label{momentooos}
\end{eqnarray}
Because of the simplifying halo axial symmetry around $OZ$, $J_1=J_2$. We can then write the integrand of Eq.~(\ref{integrando}) in terms of Eq.~(\ref{momentooos}) as
\begin{eqnarray}
    I_{20} &=& \sqrt{\frac{4 \pi}{5}} \int dV \: \rho \cdot [2(z^2+x^2)-2(y^2+x^2)+(y^2-x^2)]  
    \nonumber \\ 
    &=& -\sqrt{\frac{4 \pi}{5}} (J_3-J_2)
    \label{inercia}
\end{eqnarray}
This relation allows a chain of functions from the velocity to the potential to $J$, in brief $V(\Phi[J(\rho(r))]) $.

The data on $V(r)$ can be used to fit the density functions $\rho_0$, $\rho_2$ and $\rho_4$
by the minimum square method; from each $(R_i,\rho_i)$ parameter set (the number of which is counted to obtain $\chi^2/{\rm dof}$), the $J$s are calculated and from them the potential and the velocity to be fit. After Minuit converges, a pair of $J_1$ and  $J_3$ is extracted for each galaxy halo.

These $J$s can then be translated into the length of the semiaxes of an equivalent ellipsoid.
In the geometry of~Fig.~\ref{haloelongadoforma}, $b=c$ are the two equal semiaxes that yield a symmetric (nonrigid) top, with $a$ being the distinct one ($J_3-J_2<0$ or equivalently $a>b$ then corresponds to a prolate body, $a<b$ to an oblate one). 

\begin{figure}[h!]
  \centering
  \includegraphics[width=.5\linewidth]{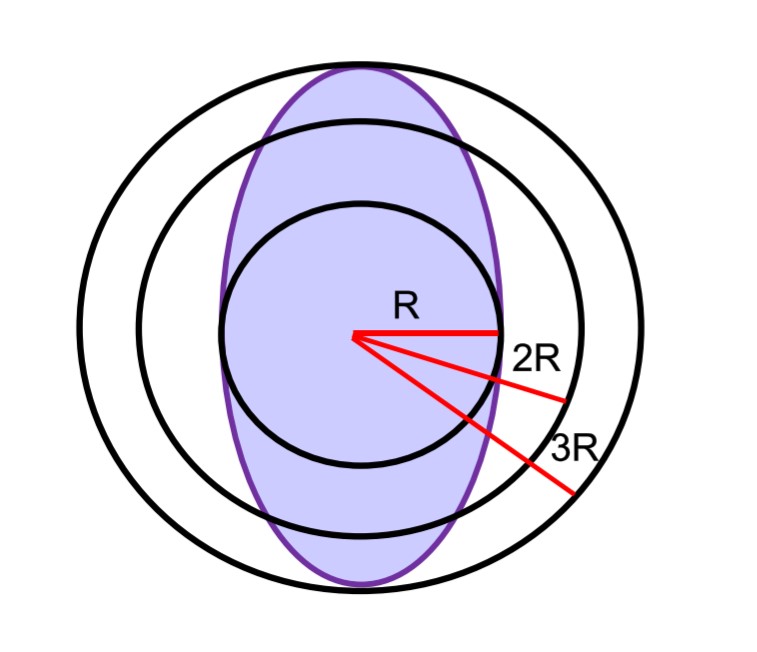}\\
  \includegraphics[width=0.7\linewidth]{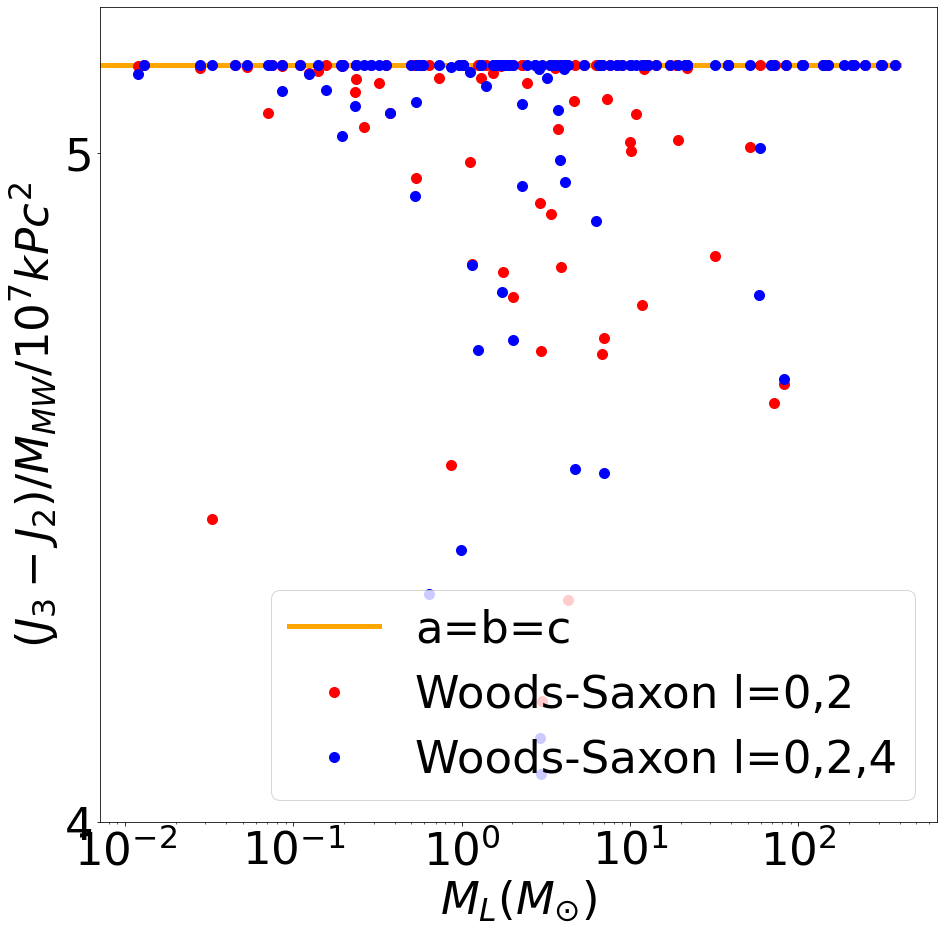}
\caption{Multipole expansion of the potential. Top: scheme for the integration over the halo. Bottom: Scatter plot of the extracted difference of the moments of inertia (normalized by the Milky Way mass). 
There is an accumulation of only slightly prolate haloes near zero; also a large number of extremely prolate ones; and very few that are oblate, and those are very nearly spherical and accumulate near $s=1$ (0 in logarithmic scale).
}
  \label{momentodeinerciaMEP}
\end{figure}

To pass from the density to $J_3-J_2$ we integrate over concentrical spherical shells with radius ranging in $r \in [0, R(\theta)]$ (see Figure~\ref{momentodeinerciaMEP}). The expression for the upper limit can be obtained from the equation for an ellipsoid
\begin{equation}
    R(\theta ) = \frac{1}{\sqrt{\frac{\sin^2{\theta}}{b^2}+\frac{\cos^2{\theta}}{a^2}}}.
\end{equation}
Comparing Eqs.~(\ref{integrando}) and (\ref{inercia}) we obtain the expression for the inertia moments
\begin{equation}
    J_3-J_2 = \frac{1}{2}\int^{2 \pi}_0 d\phi \int^{1}_{-1} d\cos{\theta} \int^{R(\theta)}_0 dr \: r^4 \:\rho_{WS}(r) (1-3 \cos^2{\theta}).
    \label{dif_inertia_momenta}
\end{equation}
Because $a$ is not known a priori for each galaxy, it is iteratively increased until the fit to $V(r)$ has converged.

Upon examining  the values of the moments of inertia, the obtained $J_3-J_2$ turns out to take negative values (see Table~\ref{i3i2} and bottom scatter plot of Fig.~\ref{momentodeinerciaMEP}), meaning that $a>b\simeq c$ and therefore the halo has a prolate shape. \\

From section~\ref{seccionpotencial} we can extract the difference between the inertia moments $J_3-J_2$ from the quadrupolar term of the potential in Eq.~(\ref{dif_inertia_momenta}).

\begin{table}[h!]
    \begin{tabular}{|c|c|c|c|}\hline 
        Order  & $\bar{X}(\sigma)\times 10^{6}  $ & Median & Geometric mean\\ \hline 
        $l=0$, 2     & $-1.5\pm 5.4$ &  $\lesssim 0 $ & $-1.50\times 10^{-6} $
\\\hline 
        0, 2, 4  & $-1.8 \pm 6.6$ &  $\lesssim 0 $ & $-1.77\times 10^{-6}$ \\ \hline 
    \end{tabular}
    \caption{ Difference of moments of inertia $\frac{J_3-J_2}{M_{VL} MPc^2}$: statistical parameters of the distribution in figure~\ref{momentodeinerciaMEP}. About a quarter of the galaxies are very prolate, the rest of them cluster near spherical shape.}
    \label{i3i2}
\end{table}

\section{Multipole expansion of the DM density}
\label{secciondensidad}

In this section, to explore the systematics, instead of directly performing a multipolar expansion of the potential, we expand the dark matter density and afterwards calculate the potential.
The method followed is akin to the one used in nuclear physics for the study of collective vibrational shell models \cite{shellmodel1, shellmodel2} where it is used  to study the deformation of certain nuclei and also heavy ion collisions \cite{shellmodel3}.

We start again from a Woods-Saxon density as in Eq.~(\ref{saxonformula})
\begin{equation}
\rho_{W-S} (r) = \frac{\rho_0}{1+e^{(r-R(\theta,\varphi))/a_0}}\ ,
\end{equation}
the difference being now that the central position of the halo edge
$R$ depends on the angular visual around the galactic center, unlike in Eq.~(\ref{saxonformula}) where it was constant.
We expand this $R(\theta,\varphi)$   in terms of spherical harmonics,
noticing that with $m=0$ they are real, and the $\beta$ coefficients are also real numbers,
\begin{equation}
    R(\theta, \phi) = R_0 \cdot \left[1+\sum^{\infty}_{l=1}  
    \beta_{l0} Y_{l0}(\theta, \phi)\right] \ .
    \label{ecuacionR}
\end{equation}
We now apply the same symmetry conditions as in section~\ref{seccionpotencial}, 
and the only terms left will then be those with even $l$ and $m=0$. 
We will once more truncate the expansion including only the monopolar, quadrupolar and hexadecapolar terms, leaving two $\beta$ deformation parameters:
\begin{equation}
    R(\theta, \phi) = R_0 \cdot \left[1+ \beta_{20} Y_{20}(\theta, \phi)+\beta_{40} Y_{40}(\theta, \phi)\right]
\end{equation}
The first one, $\beta_{20}$, is related to the elongation of the ellipsoid of revolution that represents the shape of the halo, as can be seen for a solid body (the limit in which the Woods-Saxon edge is set to zero) for which it takes the form
\begin{equation}
    \beta_{20} = \sqrt{\frac{16 \pi}{45}}\frac{a^2-b^2}{a^2+b^2} \ .
    \label{eq:semiejes_long}
\end{equation}
When $\beta_{20}>0$ the ellipsoid is prolate, becoming oblate when $\beta_{20}<0$. According to Eq.~(\ref{sphericalshelling})  the gravitational potential is then calculable as
\begin{equation*} 
     \Phi(\textbf{r})=\frac{-2\pi G}{a\cdot r}\int^\pi_0 d\theta'\int^{r/a}_0 du' \: \frac{bu^2}{1+e^{u-R/a}}+
\end{equation*}
\begin{equation}
     -2\pi G\int^\pi_0 d\theta'\int^{R/a}_{r/a} du' \: \frac{bu^2}{1+e^{u-R/a}}
     \label{eq:densityexpansion}
\end{equation}

\begin{figure}
    \centering
    \includegraphics[scale = 0.4]{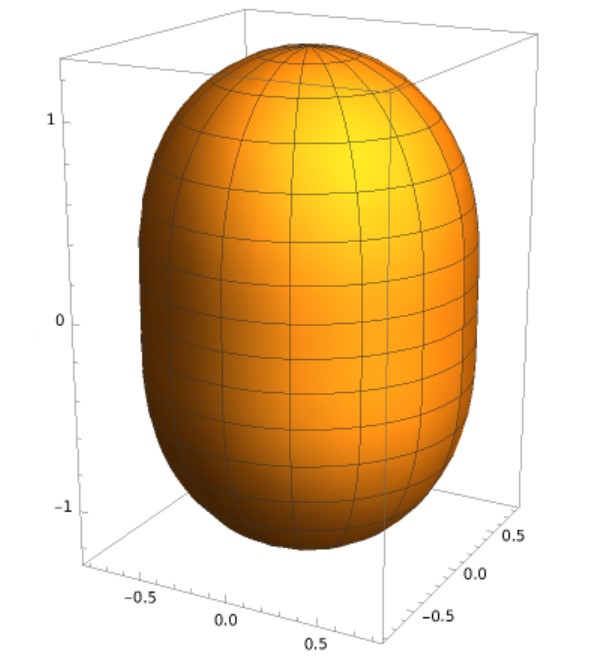}
    \caption{Halo shape obtained from the multipole expansion of the density distribution in Eq.~(\ref{ecuacionR}) for the parameters $\beta_{20}=0.7$, $\beta_{40}=0.2$, $\beta_{60}=0.1$, $\beta_{80}=0.1$; larger values of the higher order coefficients lead to unnatural oscillations of the edge, reflecting those of the Legendre polynomials, that we avoid by restricting their maximum values in the fitting.
    \label{ExpDensDraw}}
\end{figure}

We can expand Eq.~(\ref{ecuacionR}) to higher orders, and have done so for exploratory purposes. 
But for large values of higher $\beta$ coefficients (positive when prolate and negative when oblate), the surface of the halo can oscillate due to the intrincate shape of the Legendre polynomials. To avoid it we need to impose a condition of convexity, best expressed in Cartesian coordinates as $z''(x)<0$ (because the figure has azimuthal symmetry, $x=r_\perp$ can be any direction in the $XY$ plane).
This restriction binds the values that the  parameters $\beta_{l0}$ can take as $|\beta_{20}| \leq 0.7$, $|\beta_{40}| \leq 0.2$, $|\beta_{60}| \leq 0.1$, and  $|\beta_{80}|\leq 0.1$. Within these bounds the dark matter halo shape is a reasonable elongated body interpolating between a sphere and a filament, as shown in figure~\ref{ExpDensDraw}.

The fit with the most parameters is  that with $l=0,2,4$ (that depends on $\rho_0$, the central density, $a$, the skin-thickness parameter, $R_0$, the halo width in the equatorial plane, and the intensities $\beta_2$ and $\beta_4$ for the higher multipoles). The number of galaxies that have at least five data points so that a $\chi^2/{\rm dof}$ makes sense is 165; the other 10 galaxies in SPARC files are discarded for this fit.  We must also discard galaxies with unacceptable rotation curves (37 galaxies altogether for this exercise, see the explanation in Table~\ref{galaxias_descartadas}). Finally, we perform the fits for the remaining 129 galaxies of SPARC's database.

We have seen that acceptable values for the coefficients are $\beta_2 \in [0, 0.7], \beta_4 \in [0, 0.2]$ (condition of convexity). We will test a wider parameter region to study the overall behaviour of the fits. (This has only been done for fits with $l=0,2$ due to the time consuming calculations). We find that a typical fit quality with $l=0,2,4$ is not significantly better than a fit with only $l=0,2$, see as an example Fig~\ref{fig:rotNGC0300dens}. For many applications, keeping only the quadrupole term will be sufficient. In this case we study four parametrizations; therefore the ranking values range from 1 to 5, with 1 being the best fit. As one might expect, the models which fit better are those in which we leave more freedom in $\beta_2$ (see Table~\ref{modelostable_expdens}). Since the acceptable values for $\beta_2$ are constrained to a small region and Python's {\tt iminuit} gets stuck in local minima easily, the results might confuse us. 
It is then relevant to study the dependence of the fits for $\beta_2$, to see if they naturally assign it physical acceptable values (respecting the condition of convexity). 

\begin{table}[h!]
    \centering
        \caption{We attempt to order the several fits with various angular-dependence parametrizations and parameter ranges for $\beta_2$ and $\beta_4$. Given are the number of fit parameters, the mean value of the $\chi^2/{\rm d.o.f}$, its median value and its standard deviation. We find quite interesting that allowing $\beta_2<0$ does not improve the $\chi^2$. When the values of $\beta_2$ and $\beta_4$ are allowed to exceed those that yield a convex halo, the fit improves but of course the halo is probably not very physical. As is natural, the fit with $l=0,2,4$ is better than that for $l=0,2$ with equal parameter ranges.}
    \label{modelostable_expdens}
    \begin{tabular}{|c|c|c|c|c|c|}\hline 
      \textbf{Angular} & \multicolumn{2}{c}{Parameter limits}   &  $ $       &  \textbf{MED}     & \\
               {shape} & \textbf{$\beta_2$}& \textbf{$\beta_4$} & $N_{F.P.}$ & \textbf{$\pm$MAD} & $\bar{R_i} \pm \sigma$ \\ \hline 
        
        $l=0$, 2    & [0, 3]       & -        & 4 & 2.0$\pm$1.0 & $2.6 \pm 1.2$ \\\hline
        $l=0$, 2    & [0, 10]      & -        & 4 & 3.0$\pm$1.1 & $2.7 \pm 1.3$ \\
        $l=0$, 2, 4 & [0, 0.7]     & [0, 0.2] & 5 & 3.0$\pm$1.0 & $2.9 \pm 1.7$ \\
        $l=0$, 2    & [-3, 3]      & -        & 4 & 3.0$\pm$1.2 & $3.1 \pm 1.4$ \\\hline 
        $l=0$, 2    & [0, 0.7]     & -        & 4 & 4.0$\pm$1.3 & $3.7 \pm 1.2$ \\\hline 
        
    \end{tabular}
\end{table}

\begin{figure}
  \centering
  \includegraphics[width=\columnwidth]{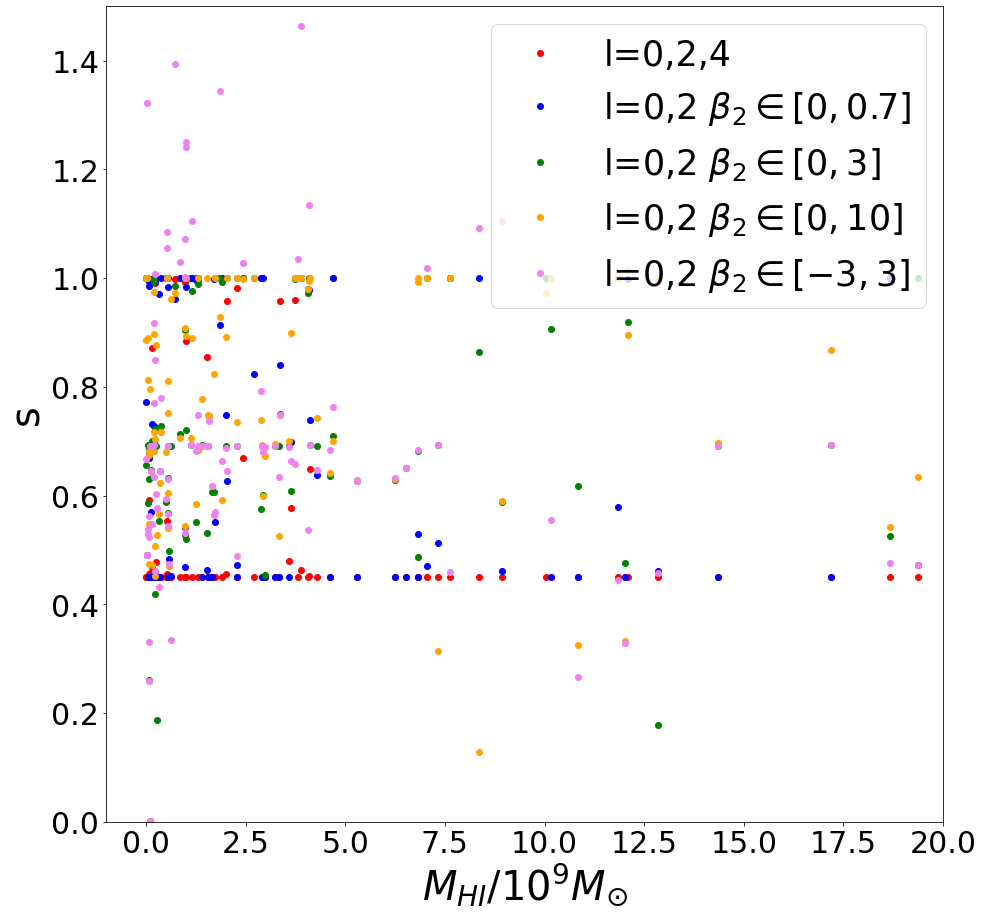}
  \caption{Ratio between the semiminor axes $b=c$ and the semimajor axis $a$ for haloes with ellipticity that ranges within $s=c/a \in(0, 1)$. The haloes that in an unconstrained fit would turn up to be oblate are clustering at the line $s=1$. The ones clustering around $s=0.45$ would probably want even larger prolateness but we do not allow this due to the halo shape becoming unnaturally wavy for the parametrization with few angular multipoles and large $\beta_{20}$.  }
  \label{axesrelation}
\end{figure}

\begin{figure}[h!]
    \centerline{\includegraphics[width=0.4\textwidth]{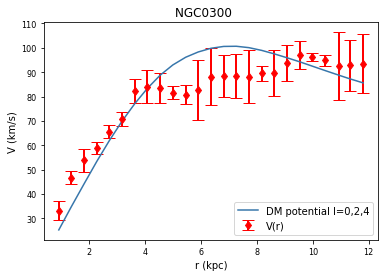}}
    \caption{We fit the rotation curve of NGC0300 using the $V\left(r\right)$ of the multipole expansion of DM density distribution, Eq.(\ref{eq:densityexpansion}). SPARC data from~\cite{sparc}.}
    \label{fig:rotNGC0300dens}
\end{figure}

In Fig.~\ref{axesrelation} we used Eq.(\ref{eq:semiejes_long}) to obtain the relation between the major and minor axes. In four of the fits we allowed $\beta_2$ to only take positive values. The triaxiality is taken as zero as in the rest of the manuscript. These haloes that we fit in the end, will range from spherical $s=c/a=1$ ($\beta_2\simeq0$) to very prolate haloes $s=c/a=0$ ($\beta_2>0.7$). In Fig.~\ref{axesrelation} we see that for these models  the trend is for the halo to become prolate rather than spherical. There is a fifth fit in which we allow $\beta_2$ to take negative values. In Fig.~\ref{axesrelation} we see that there are not many haloes that are preferably oblate. 

We find that, in general, haloes come out of this fit neither extremely prolate nor nearly spherical. Those models with a best fit $\beta_2\in[0, 3]$ and $\beta_2\in[0, 10]$ tend to give prolate haloes, with values of s ranging between (0.45, 1.00). We also realise that, when giving $\beta_2$ freedom to take negative values, the number of haloes that are preferably oblate is relatively small.

\section{Statistical extraction of the halo ellipticity} \label{extract_s}

We now turn to the statistical characterization of the ellipticity $s$ over the sample.

\subsection{ Definite model: Exponential Ellipsoid Parameterization}
\label{subsec:expellipsoid}

In this subsection we report a model-dependent extraction of the ellipticity from a simple, generic exponential density distribution. We choose it to have its surfaces of constant density be ellipsoids with revolution symmetry ($b=c$, that is, we do not consider triaxiality that does not play a role in $V(r)$) satisfying 
\begin{equation}
\frac{x^2+y^2}{c^2} + \frac{z^2}{a^2} = \rm const
\end{equation}
for which the ellipticity is trivial to assess. In cylindrical coordinates, with $r=r_{\perp}\equiv\sqrt{x^{2} + y^{2}}$, 
an apt choice is then
\begin{equation} \label{ellipsoid0}
   \rho\left(r,z\right)=\rho_{0}e^{-\frac{1}{b}\sqrt{ r^{2} + \left(b/c\right)^{2} z^{2} }} \ .
\end{equation}
Introducing once more the ellipticity parameter $s=c/a$ and redefining $R:=b$ we have
\begin{equation} \label{ellipsoid}
   \rho\left(r,z\right)=\rho_{0}e^{-\frac{1}{R}\sqrt{ r^{2} + s^{2} z^{2} }}\ .
\end{equation}
This density profile has four degrees of freedom: the ``hidden'' power law of the argument of the exponential, which is fixed to 1, and the three manifestly free parameters $\rho_{0}$, $R$ and $s$.

There is no analytical expression for the gravitational field of this density profile, and thus we calculate it numerically from the general expression
\begin{equation}
    \mathbf{g}\left(\mathbf{
    x}\right)=-G\int d^{3} x{'}\rho\left(\mathbf{x{'}}\right)\frac{\mathbf{x}-\mathbf{x{'}}}{\left|\mathbf{x}-\mathbf{x{'}}\right|^3}\ .
\end{equation}

Since the spiral-galaxy rotation curve $V(r_\perp)$ is measured on the galactic plane, we only need the radial component $g_\perp$, which in spherical coordinates can be written as
\begin{equation}
\label{eq:ggeneral}
    g_{r}\left(r\right)=\int d^{3}x' \frac{-G\rho\left({\bf r}'\right)\cdot (r-r'\sin{\theta'}\cos{\phi'})}{\left(r^{2}+r'^{2}-2rr'\sin{\theta'}\cos{\phi'}\right)^{3/2}}
\end{equation}

\begin{figure}
\includegraphics[width=0.9\columnwidth]{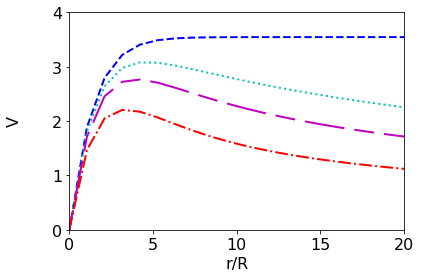}
\caption{\label{fig:Vofq} $V(r)$ rotation curve for the ellipsoidal dark matter distribution of Eq.~(\ref{ellipsoid}). The extreme values are $s=1$ (dashed-dotted line, red online) corresponding to a spherical distribution, at the bottom, and $s=0$ (dashed line, dark blue online), the limit of a cylindrical distribution, at the top.
Also given are intermediate values $s=0.4$ (dashes of alternating size, purple online) and $s=0.2$ (dotted line, light blue online). }
\end{figure}

Finally, we obtain the rotation curve through Eq.(\ref{eq:vgform}). Adequately normalized ones are plotted in Fig.~\ref{fig:Vofq} for different values of $s$. The dashed line at the bottom corresponds to $s=1$: the source is spherically symmetric and $V\left(r\right)$ decreases for $r/R>5$ as per Kepler's $3^{\rm rd}$ law. As we give $s$ smaller values,   $V\left(r\right)$ becomes less slanted, 
and when $s=0$ (the top curve) we see the characteristic flattening of cylindrical sources.
Thus, we have an appropriate interpolating model between cylindrical and spherical geometries as function of one parameter.

By fitting the same subset of rotation curves used in Section~\ref{sec:modelprofiles} we can obtain the value of $s$ as well as its 1--$\sigma$ uncertainty. Minimization with {\tt iminuit} 
becomes slow due to the triple integration in Eq.~(\ref{eq:ggeneral}). 
To reduce running time it is convenient to use an adapted limit of integration for the radial variable, $r'\in \left[0,r+10R\right]$. 
We are satisfied with relative numerical errors below the 10\% level that are under the typical
statistical uncertainty in the $s$ ellipticity parameter: 
$\delta_{V}=\left(V_{\rm E}-V_{\rm N}\right)/V_{\rm E}$ reaches its maximum for $s=0$ (cylindrical source) as $\delta_{V}=0.05$ for $r=10R$ and $\delta_{V}=0.09$ for $r=20R$, increasing with radius.

In Fig.~\ref{fig:s} we show a histogram of $\log{s}$, based on the logarithmic scatterplot of Fig.~\ref{fig:selip} with the $s$ values obtained for each individual galaxy.

 A supermajority of galaxy DM haloes is then prolate, and lies below the horizontal bar (blue online) at $s=1$ in the figure. There are however 29 galaxies with $s>1$ at 1--$\sigma$ level.

The resulting 164-galaxy population's ``central'' value for the ellipticity variable $s$  and its uncertainty $\Delta s$, within this exponential parametrization of the radial dependence, is obtained by minimising the statistical estimator
\begin{equation}
\label{eq:chi2_3}
	\chi^{2}=\sum_{i=1}^{N}\frac{\left[s^{\rm fit}_{i}-s\right]^{2}}{\Delta s^{\rm fit\ 2}_{i}}\ .
\end{equation}
This should not be applied blindly. Careful analysis shows that a very few galaxies with very small uncertainty band in their own ellipticity have a disproportionate effect on the central value of the entire set. This may be a reason why earlier literature found contradictory results.

For example, UGC02916 tends to prefer an oblate halo and seems to yield an amazingly precise ellipticity of $s=2.63398(1) $.
Due to such unbelievable goodness of fit, this one galaxy almost by itself pulls the central ellipticity to $\bar{s}\pm \Delta s= 2.14901(1) $, on the oblate side as it is larger than 1.

\begin{figure}[h!]
\centering
\includegraphics[width=0.5\textwidth]{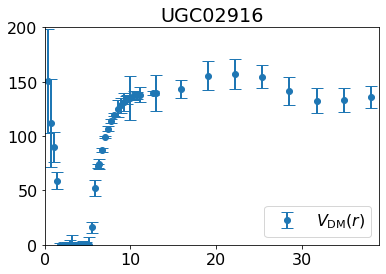}
\caption{(Pseudo-) data $V_{\rm DM}$ after subtracting SPARC's visible matter component off the UGC02916 rotation curve. This galaxy is an example where an oblate DM distribution is favored by the fit, with an incredibly small $\chi^2$ that pulls the global fit of the entire database. But notice the pronounced minimum at few kpc that makes us doubt that the SPARC extraction of the visible matter components, leaving this squared velocity to be explained by dark matter, is totally reliable. Notice in particular that at large radii $V(r)$ flattens to a constant, as typical spiral galaxies do, suggesting prolateness in the end. It is the small $r$ part, very impacted by the visible matter, that is driving the fit.}
\label{fig:gal_osc}
\end{figure}

\begin{figure}[h]
\centering
\includegraphics[width=0.4\textwidth]{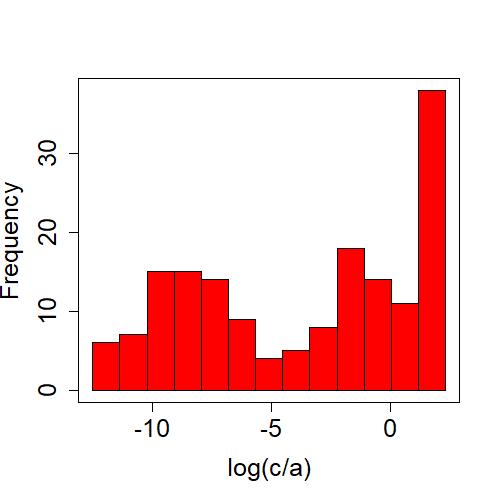}
\caption{ Histogram of the distribution of $\log{c/a}$ for the exponential ellipsoid parameterization. It is clear that the data suggests many very prolate DM haloes with negative log, whereas those that come out oblate are not very much so, but stay relatively near the spherical shape.}
\label{fig:s}
\end{figure}

\begin{figure}[h!]
\centering
\includegraphics[width=1\textwidth]{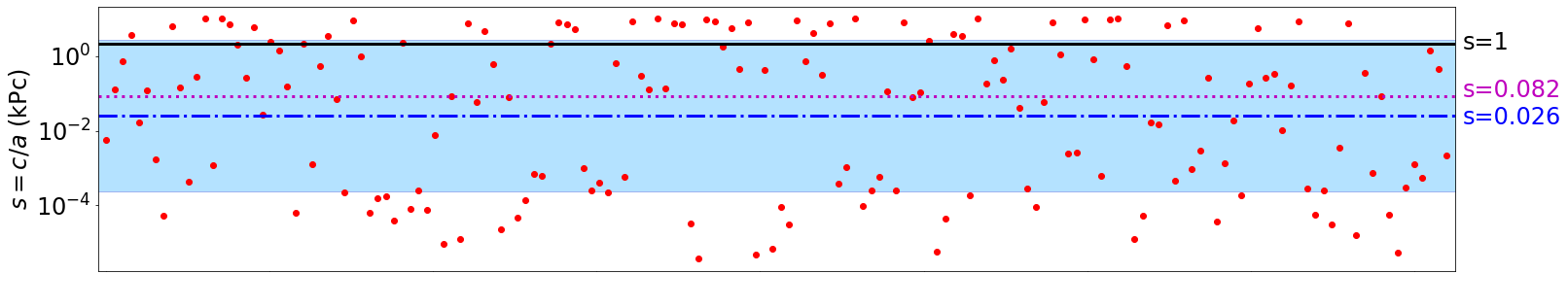}
\caption{Ellipticity ratio $s=c/a$ obtained from fitting an exponential ellipsoid parameterization to the subset of rotation curves  from the SPARC database selected in section \ref{subsec:expellipsoid}\cite{sparc}. 
Most galaxies lie underneath the $s=1$ solid line (black online), suggesting prolate DM haloes. The dotted line (purple online) denotes the median of the distribution, while the dashed line and the shadowed region (blue online) denotes the geometric mean and its confidence interval.}
\label{fig:selip}
\end{figure}

Closer examination of this one galaxy (see Fig.~\ref{fig:gal_osc}) shows that at large $r$ the behavior of $V(r)$ is that of a typical spiral galaxy with a prolate halo, and that 
the $s>1$ value is driven by the small-$r$ dip, very suggestive of a poorly controlled visible-matter distribution. 
\color{black}

We have individually examined the minority of galaxies that have positive $s-1$ and suggest an oblate DM halo. Most of them have a rotation curve that is {\it increasing} with distance out to the farthest measured point, that is, measurements have probably not extended far out enough to see the typical settling into a flat rotation curve. 
A much smaller number of others fall in alternative categories (large uncertainties, very untypical behaviours such as a quick falling of $V(r)$ as a step function, etc.) and are classified in Appendix~\ref{app:oblate}.

 This apparently outlier galaxy, UGC02916,  belongs to the first class of the categories there defined: its pseudo-data DM rotation component exhibits a pronounced oscillation at small $r$. Excluding this one galaxy from the analysis we immediately obtain an extremely prolate (and unreasonably accurate) value for the distribution, $s=0.05845(2)$. Plucking off the next-to extreme galaxy, in this case a prolate one, returns the central ellipticity to order $s\simeq 0.3$; this instability of the global fit to a few galaxies is typical
 of statistical samples with outliers, and the correct procedure is to remove them. 
 
 Since there is some ambiguity in the point where the remotion of outliers needs to stop (several conventions are used in the literature), we have opted for iterating the process of removing one value at a time with lower uncertainty in $s$ to obtain a sequence of central $s$ values for a decreasing number of galaxies $s_n$.

We consider that the value computed is reliable when two consecutive values of $s$ are compatible at 1--$\sigma$. This first happens after removing 32 galaxies, leading to a temptative result at $1\sigma$ of $s=0.693\pm 0.027$. 

It is in carrying out this exercise when we have realized that the more revealing statistic estimator is the geometric mean, or the logarithmic average, 
as discussed below in subsection~\ref{subsec:comparison}.
Indeed, if the geometric mean of each galaxy's ellipticity is taken, 
$\sqrt[n]{s_1\cdot s_2 \dots s_n}$, it turns out to yield a very prolate value, $\exp{\langle \log s\rangle}=0.026$
which amounts to the longer axis being 38 times larger than the shorter ones !`on average! However, the spread is large, with a confidence interval for the variable $\log s$ that is $\log s = -3.6 \pm 4.7$ 
or, exponentiating, $s = [0.0002,  2.783]$. That is, there are some  DM haloes that are preferably fit to truly filamentary shapes with $ s \ll 1$, while a few can be somewhat oblate, like a thick pancake. On log average, it is clear that they tend to strong prolateness.

\onecolumngrid

\twocolumngrid

\newpage
$ $

\newpage

$ $

\newpage
\subsection{Extraction from the multipole expansion of the DM density} \label{subsec_s_extraction2}

In this subsection we quickly turn to the extraction of the ellipticity from the multipole expansion in section~\ref{secciondensidad}. There, we already advanced, in Figure~\ref{axesrelation} the ratio of the minor to major semiaxes.

We here discuss the statistical distribution of that ellipticity $s=c/a$, after having noted that the 
average of the logarithm of this ratio is very important. This should be similar to using the geometric mean instead of the arithmetic mean of the distribution of $s$.

\begin{figure}[h]
\centering
\includegraphics[width=0.4\textwidth]{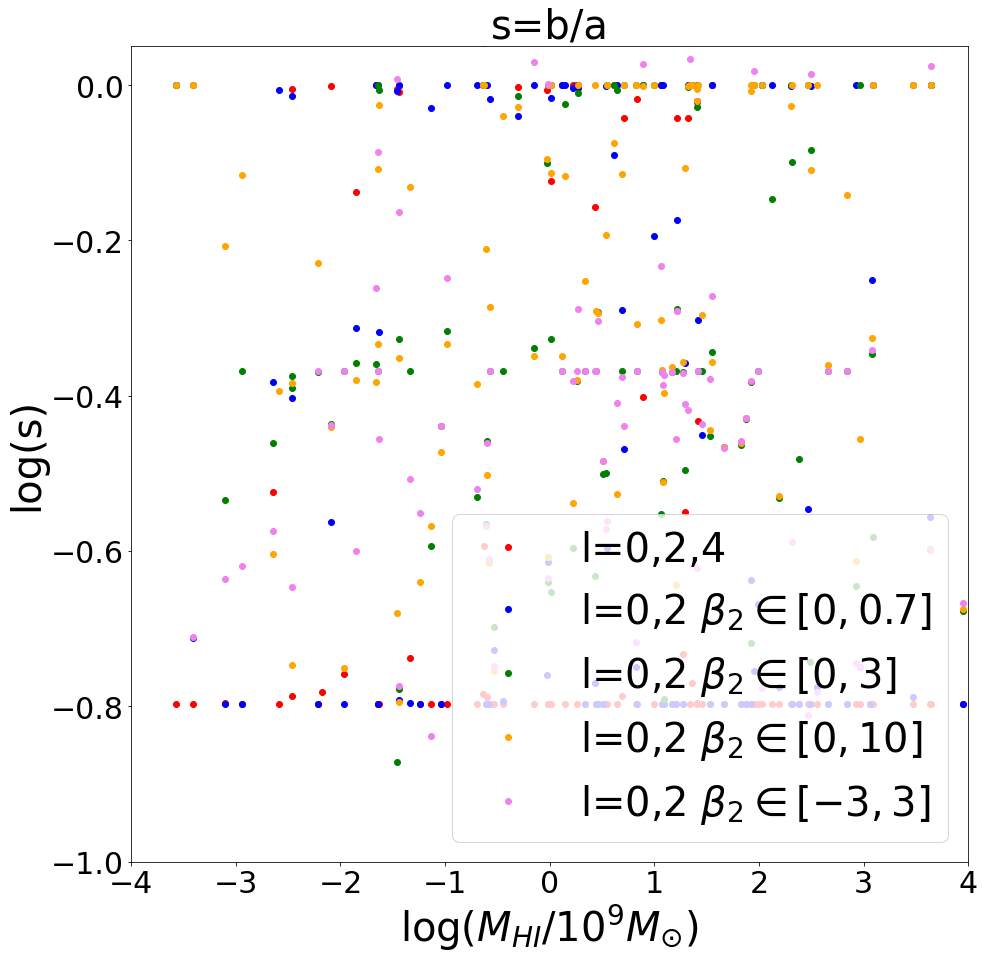}
\caption{Data from Figure~\ref{axesrelation} above, replot in log scale.}
\label{fig:s_log}
\end{figure}

The data from Figure~\ref{axesrelation} are then replotted in Figure~\ref{fig:s_log} in a logarithmic scale. 

We still see that the values of $s$ largely  fall in the region $s\in [0.45, 1]$ in spite of the wider parameter range allowed for $\beta_2$ that included negative (oblate) haloes. In addition, we realise that, after ordering the galaxies with increasing $s$, 
we find that more than  50$\%$ of them have $s\lesssim 0.6$ (see Tables~\ref{tab:resumenellipticity} and ~\ref{tab:quartiles}).   

\begin{table}[h!]
    \begin{tabular}{|c|c|c|c|c|}\hline 
       Order  & $\beta_2$ & $\bar{s}\pm \sigma$ & ${\rm MED}(s)\pm$MAD  & Geometric mean  \\ \hline
    $l=0$,2   & [0, 10]  & $0.76\pm 0.2$ & 0.75$\pm$0.17 & 0.68    \\\hline 
    $l=0$,2   & [0, 3]   & $0.72\pm 0.2$ & 0.69$\pm$0.19 & 0.65    \\\hline 
    $l=0$,2   & [-3, 3] & $0.85\pm 0.9$ & 0.69$\pm$0.39 & 0.68    \\\hline 
    $l=0$,2   & [0, 0.7] & $0.72\pm 0.3$ & 0.73$\pm$0.25 & 0.65    \\\hline 
    $l=0$,2,4 & [0, 0.7] & $0.56\pm 0.2$ & 0.45$\pm$0.16 & 0.53    \\\hline 
    \end{tabular}
    \caption{\label{tab:resumenellipticity} Multipole Expansion of $\rho$ with coefficients $\beta_2$, $\beta_4$ limited by convexity of the halo. Angular-dependence parametrization, $\beta_2$ parameter region, mean value, standard deviation, median value and geometric mean of the ratio of the length of the semiaxes s=c/a.}
\end{table}

\begin{table}[h!]
    \begin{tabular}{|c|c|c|c|c|c|}\hline 
       Order  & $\beta_2$ &  25$\%$ & 50$\%$ &75$\%$ & 100$\%$ \\ \hline
    $l=0$,2   & [0, 10]  &  0.60 & 0.74 & 0.99 & 0.99  \\\hline 
    $l=0$,2   & [0, 3]   & 0.58 & 0.69 & 0.99 & 0.99  \\\hline 
    $l=0$,2   & [-3, 3] &  0.54 & 0.69 & 1.00 & 9.3 \\\hline 
    $l=0$,2   & [0, 0.7] &  0.45 & 0.73 & 0.99 & 0.99 \\\hline 
    $l=0$,2,4 & [0, 0.7] &  0.45 & 0.45 & 0.48 & 0.99  \\\hline 
    \end{tabular}
    \caption{\label{tab:quartiles}
    Quartiles of the distribution of values obtained for the ellipticity ratio s. From left to right, the columns show the angular-dependence parametrization, the $\beta_2$ parameter interval, and the maximum value of $s$ reached within each quartile of the $s$ distribution. Because we have taken a curated galaxy sample paying attention to eliminating those with steep oscillations, three quarters of all galaxies are clearly prolate rather than oblate (in the bottom row, the percentage of prolate ones is even larger).}
\end{table}

The outcome of this last fit is that three quarters of the sampled galaxies prefer prolate dark matter haloes, and both median and geometric mean suggest typical ellipticities $s\sim 0.6-0.7$, whereas in this case, the arithmetic mean is less clear.

\section{Concluding discussion} \label{sec:discussion}
We have reported fits to the galactic rotation curves obtained from SPARC's database \cite{sparc} using 
several approaches.
Because of the different number of parameters of each, a varying number of galaxies with few measured data points have been left out in each, in line with prior work~\cite{Loizeau:2021bum}.
Additionally, in the analysis of sections~ \ref{secciondensidad} and \ref{seccionpotencial}  we have left out the galaxies specified in table~\ref{galaxias_descartadas} of the appendix below, that had violent oscillations in the rotation curve. In other analysis we have, however, included them, to avoid introducing too much overall bias. 

From SPARC's own analysis~\cite{lelli} we have taken the basic data (distance to the galactic center $r$, total velocity $v\pm e$) but also the separate square velocity contributions for each component of visible matter. These comprise the bulge of the galaxy, its disk, and gas cloud. 
We have adopted their relation for the mass-to-light proportions $\Upsilon_{bulge} = 1.4\Upsilon_{disk}$ and $\Upsilon_{disk}=0.5 M_\odot/L_\odot$~\cite{lelli,lellithesis} to be able to subtract the barionic contribution to the rotation curve.
\begin{equation}
    V_{{\rm B}, i} = \sqrt{V_{i, \rm gas}^2+\Upsilon_{\rm bulge} V_{i, \rm bulge}^2+ \Upsilon_{\rm disk}V_{i, \rm disk}^2}
    \label{barionica}
\end{equation}
(with $i$ labelling each galaxy) and our actual fits refer to the rest,  $V_{\rm DM}$, presumably due to the dark matter distribution, with the total velocity as in
\begin{equation}
    V = \sqrt{V_{DM}^2+V_{B}^2} \ .
    \label{ajustevelocidad}
\end{equation}
Different approaches were contrasted against an adequately defined $\chi^2$ function~ \cite{estadistica,iminuit},
\begin{equation}
    \chi^2 = \sum^N_{i=1}\frac{(V_{\rm th}-V_{\rm obs})^2}{e_{\rm obs}^2}\ .
\end{equation}
We normalized this per degree of freedom, $\chi^2/(N-k)$ by dividing through the difference between the number $N$  of data points for each galaxy and the number of fit parameters $k$.

Farrar and Loizeau also performed analogous fits to the rotation curves in~\cite{Loizeau:2021bum}; further models can be seen there.
Their generic conclusion is that the Einasto profile, or also even an additional disk component, would provide a better fit than traditional dark matter models or self-interacting dark matter.

A main feature of our allowing the use of elongated shapes is that the flatness of the rotation curves is more natural, leaving much more freedom to the underlying dark matter profiles as function of the distance, that do not need to be perfectly isothermal and therefore the underlying microscopic physics~\cite{Chavanis:2021jds} is less constrained.
Overall, we do find that fits with prolate haloes are preferred for non fine-tuned radial dark matter density profiles (in those that approach the precise $\rho\sim 1/r^2$ power-law form there is quite some degeneracy in describing $V(r)$ for large $r$ and the shape cannot reliably be extracted).

\subsection{Comparison with large-scale numerical simulations and other work.}
\label{subsec:comparison}

In the simulations reported by Allgood {\it et al.} and other
works~\cite{Allgood:2005eu,Bonamigo:2014rba,Vega-ferrero:2016xoa,Ceverino:2015oaa}, 
the dark matter halo distribution was found to be slightly triaxial,  
and slightly more prolate than oblate, with a mean ellipticity compatible with our typical values. 
However, those authors do not seem to stress the point that a few of the galaxies that are very oblate 
are actually pulling the fit towards oblateness, when a significant majority of them is actually prolate, 
some being extremely prolate, and this providing what should become a textbook explanation for the flatness 
of rotation curves.

In fact, the procedure of quoting an arithmetic average of the ellipticity $a/c$ is obscuring the actual 
stand of the galaxy population.
A quick way to see it is that the average of two numbers, 0.33333 and 3, the first of which is as prolate 
as the second is oblate, becomes $1.666>1$ which is clearly oblate. However this population of two galaxies 
should be neutral and yield an average spherical shape, with $\langle s\rangle= \langle a/c \rangle =1$. 
Obviously, the correct averaging procedure for a variable distributed
over $(0,\infty)$ with neutral point at $s=1$  is to work in a logarithmic scale.

Therefore we propose to average the natural logarithm $\langle log(s) \rangle $ 
(this is equivalent to using the geometric mean instead of the arithmetic mean of the distribution of $s$) 
over the galaxy sample, and have obtained, for example in section~\ref{subsec:expellipsoid},
\begin{equation}
\langle \log(s) \rangle = 0.026
\end{equation}
which is distinctly smaller than unity. This indicates a rather prolate distribution of galaxies, 
though with a broad shoulder of some oblate ones as indicated by a spread $[0.0002,  2.783]$. 
A second, independently coded analysis based on a spherical-harmonic expansion, yields $s\simeq 0.6\pm 0.20$.

Table~\ref{tab:wrapup} presents these and other shape analysis, 
including one based on weak lensing that we have located in the literature~\cite{Hoekstra:2003vn}.
Apparently the lensing data is also suggestive of an average prolate halo, 
and once more the authors seem to be using an arithmetic average over $s$.

\begin{table}
\caption{Ellipticities of dark matter haloes extracted by various methods;
and approximate values in the logarithmic scale that we advocate. If known for several
halo masses, we quote those for large haloes that can host typical spiral galaxies.\label{tab:wrapup}}
\begin{tabular}{|c|cc|c|}\hline
Method  & $s$           & $\log s$ & Reference \\ \hline
Weak lensing & $0.66(0.07)$ &  $-0.41(0.11)$  &  \cite{Hoekstra:2003vn} \\
\hline
Fit galactic $V(r)$ &  & $-3.6(4.7)$ & This work (sec.~\ref{subsec:expellipsoid}, \\
                    &  &               & uncleaned sample) 
\\ \hline
Fit galactic $V(r)$ & $0.6(0.2)$ & $[-0.9,-0.2]$ & This work (sec.~\ref{subsec_s_extraction2},\\ 
                    &              &               & curated  sample) \\ \hline
Simulations  & & & \\
at $z=0$  & $\simeq 0.6$ & & \cite{Allgood:2005eu}\\ \hline
\end{tabular}
\end{table}

\begin{figure}   
\centering
\includegraphics[width=0.4\textwidth]{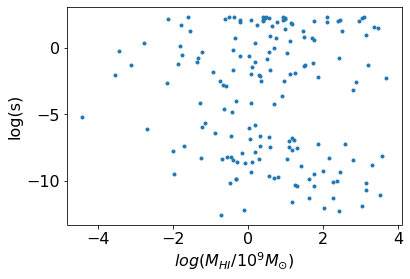}
\caption{We fail to find a visible correlation between the presumed dark-matter halo ellipticity 
and the visible galactic mass (that, by the Tully-Fisher relation, is normally taken 
proportional to the total mass).}
\label{fig:logs_logM}
\end{figure}

We quote, for the cosmological simulation entry, a number of 0.6 that broadly describes 
what is reported in figure 1 of that reference, at $z=0$. 
Those authors have also extracted the dependence with the cosmological redshift $z$ 
and with the galaxy mass $M_{\rm galaxy}$.
They see a clear correlation $s(M)$ that we cannot confirm at this point, 
as shown in figure~\ref{fig:logs_logM}.

\vspace{0.3cm}

\subsection{Final comments}

Additionally to gravitational lensing and the rotation curves, further confirmation of the shape of the halo may come from studying observables outside the galactic plane, such as stellar streams~\cite{Bonaca:2014qia}. Broadly, in the presence of such halo the movement perpendicular to the galactic rotation axis, not too far from the galactic plane, is the same as for a spherical distribution with changed parameters, and the vertical motion is that of an oscillator, with the orbital plane precessing~\cite{Llanes-Estrada:2021hnt}.

\begin{figure}[h]
\centering
\includegraphics[width=0.4\textwidth]{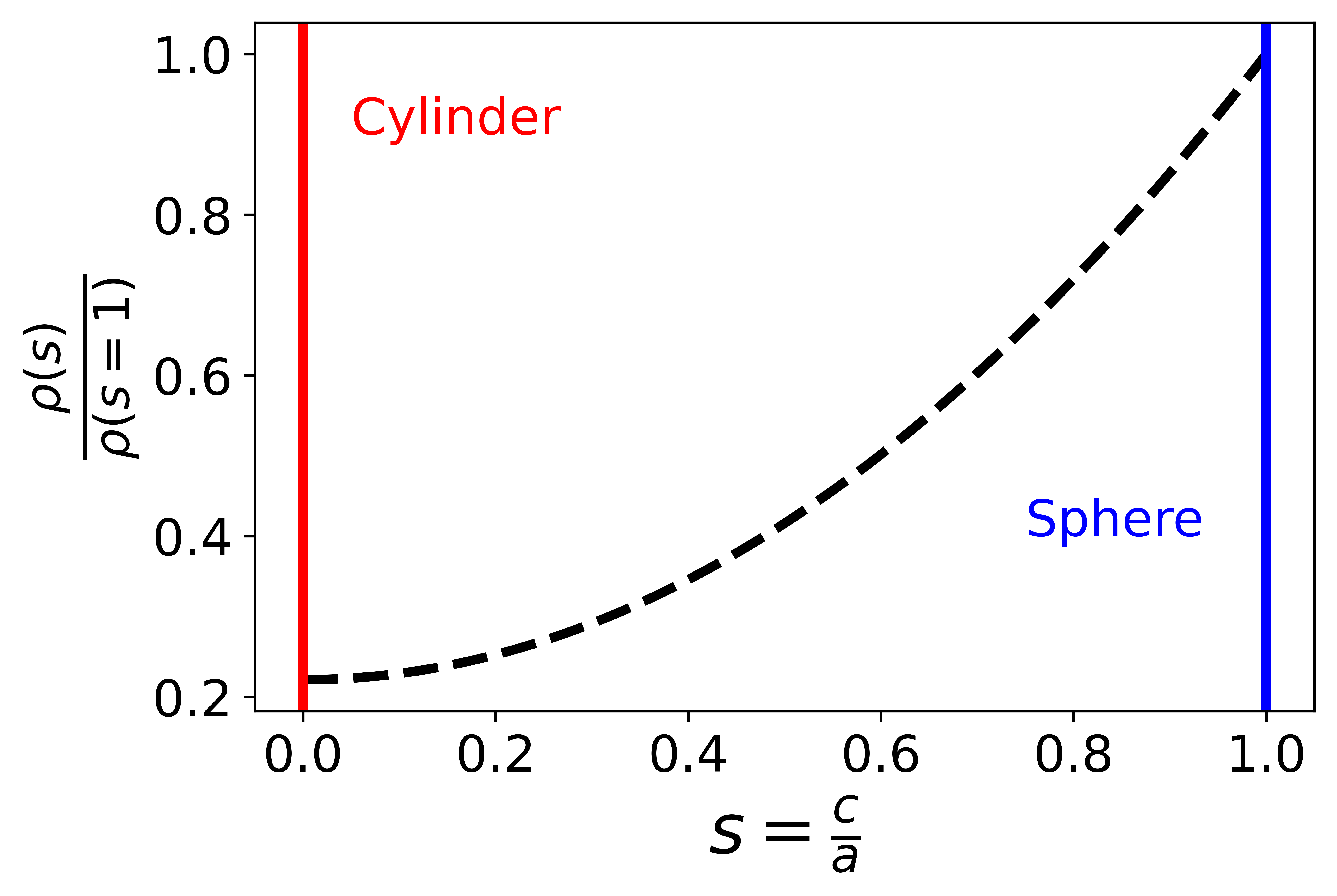}
\caption{A prolate halo implies less dark matter in the galactic plane: we show the ratio to a spherical shape, other things being equal. This would mean that extractions of DM-nucleon cross-sections in the laboratory are affected by a further factor of around 2.}
\label{fig:lessDM}
\end{figure}

These results are of impact to the direct laboratory detection programme.
As shown in figure~\ref{fig:lessDM}, as the deformation of the halo towards prolateness increases, less dark matter is to be found in the galactic disk. This entails that estimates of dark matter therein are overestimated, typically by a factor 2, which is affecting the extracted bounds on dark matter-nucleon cross-sections.

We think we have exhaustively employed the information at hand, but there is room for future improvements. For example, one roadblock that we have found is that some galaxies have rotation curves that are flat right out of $r=0$, for example NGC5371 or NGC5907; in this cases, the usual low-$r$ growth of $V(r)$ is not visible. This introduces model distortions that affect our fits; basically only an infinitely thin filament could fit that rotation curve. This  probably happens because the identification of the point $r=0$ in the galactic plane has not been fully achieved by the observational collaboration (since galaxies are most often seen at oblique angles). To reduce this uncertainty in our fits, we would need ``better'' data, which is of course beyond our ability as it depends on the observational program.

\section*{Acknowledgments and disclosure of author responsibility}

We thank our engineer David Fern\'andez Sanz for maintaining an
adequate computing environment suited to our needs at the theoretical physics departmental cluster, and C. Pieterse and N. Loizeau for useful conversations.
Financially supported by spanish grant MICINN: PID2019-108655GB-I00 (Spain),
and Univ. Complutense de Madrid under research group 910309 and the IPARCOS
institute. \\
Oliver Manzanilla Carretero is responsible for the codes and reporting of 
section~\ref{sec:modelprofiles} and subsection~\ref{subsec:expellipsoid}. Adriana Bariego Quintana wrote an independent program and reported the results of section~\ref{seccionpotencial}, \ref{secciondensidad} and~\ref{subsec_s_extraction2}. Felipe J. Llanes-Estrada designed and directed the investigation, and is responsible for the manuscript's draft. The authors have no conflict of interest.

\appendix
\section{Classification of galaxies without straightforward rotation curves}
\subsection{Galaxies with fits typically yielding oblate haloes}
\label{app:oblate}

In this appendix we first provide a classification of, and list, the galaxies that 
are seemingly actually best fit by an oblate shape than a prolate shape depending on the analysis, as they 
form a distinct minority of the SPARC database that reduces the force
of the main result of the article, so they need individual understanding.

We have found that the dark matter velocity component $V_{\rm DM}$ in Eq.~(\ref{eq:vdm}) of some of these galaxies keeps growing at large $r$. That means that the edge of their dark matter distribution has not been reached and thus we cannot really find out the shape of the DM halo.
These galaxies are assigned to class 3 in table~\ref{tab:classes}.

\onecolumngrid

\begin{table}[h!]
\caption{List of galaxies that do not favor a prolate dark matter halo shape, assigned to specific
behaviour classes as described in the text.\label{tab:classes}}
\begin{tabular}{|c|c|} \hline
Class & Galaxies \\ \hline
1 (Oscillating) & UGC02916   NGC2903 NGC2955 NGC3877 NGC4051 NGC4138     \\ \hline
2 (Uncertain) & F561-1 UGC04305 \\ \hline
3 ($V_{\rm DM}(r)$ & F568-3  F571-8  IC4202  NGC0055 NGC2903  NGC3769  NGC4157  NGC4217  NGC4389 \\ 
   ever growing) &  NGC5005  NGC5055 NGC6195  NGC7331  UGC02455   UGC06614   UGC06973  UGC07866 UGC09037   \\ \hline
 4 (small $r$) &  NGC3893 UGC05986  \\ \hline
 5 ($V_{\rm DM}(r)\simeq 0$ at large radii) &    PGC51017  UGC06628  \\
   \hline
\end{tabular}
\end{table}

\twocolumngrid

A few other galaxies have  properties that make the fitting with a prolate halo difficult, and are also listed in table~\ref{tab:classes}. 

Those in class 1 are affected by large oscillations in the $V_{\rm DM}$ pseudo-data at short distances $r$ (see Fig.\ref{fig:gal_osc}). This happens because the estimated contribution from the visible matter to the rotation curve is larger than the rotation curve itself at some points; therefore we cannot really trust the extraction of the pseudodata that has to be assigned to Dark Matter. 

The two galaxies in class 2 have very large data uncertainty, so while they come out oblate they are not very significant. 

In turn, the data for the two galaxies in class 4 reaches small distance only,
$r<4 a$ respect to the ellipsoid axis, where the dependence of the rotation curve on $s$ is smaller (see Fig.\ref{fig:Vofq}).

Finally, there are two galaxies that we assign to a  class 5 that have quite an anomalous behaviour unlike other spirals, with a rotation curve that starts off at 20 kpc and after a quick decrease to $V_{\rm DM}\sim 0$ stay there for large distances (ironically, this behavior is closer to what Kepler's law would make us expect, though the slope is too steep).

At least one of the galaxies analyzed, NGC2903, simultaneously belongs to two categories, in this case 1 and 3, because in addition to a never-decreasing rotation curve, it presents significant oscillations.  

Having achieved some understanding of why a fraction of the galaxy sample favors an oblate halo shape, we feel confident that a good explanation for the flattening of $V(r)$ for the typical spiral galaxy, the majority behaviour, 
can be the prolateness of the dark matter distribution.

\subsection{Galaxies rejected in Sections \ref{secciondensidad}, \ref{seccionpotencial}
on the grounds of strongly distorted rotation curves and similar.}
In Sections~\ref{secciondensidad}, \ref{seccionpotencial} we have rejected some of the galaxies from SPARC's database and not included them in the fit due to problems that possibly arise from the measurements of the rotation curve. The classification of those problems with $V(r)$ is provided in Table~\ref{galaxias_descartadas}. In some of the galaxies we find abrupt oscillations in the first few points (Class 2) and/or oscillations in the middle of the rotation curve (Class 1): these oscillations difficult the fitting of those rotation curves. Although this may be an observational issue, our guess is that it is due to inhomogeneities and/or asymmetries of the visible matter that is subtracted to obtain the $V_{\rm DM}(r)$ that we actually fit.
For several galaxies there are not enough data near the central region (Class 3) so that the best fit is basically provided by an infinitely thin filamentary source, so we do not employ them to extract a halo ellipticity in sections\ref{secciondensidad} and \ref{seccionpotencial} as they would considerably distort the fit favoring prolateness. 

\onecolumngrid

\begin{table}[h!]
    \begin{tabular}{|c|c|} \hline
        Class & Galaxies \\ \hline
        1 ($V(r)$ Curve oscillates &  IC4202, NGC0891, NGC2683, NGC2955, NGC3726, NGC3992,\\ with some intensity)
         & NGC4013, NGC4214, NGC5033, NGC5055, NGC5371, UGC06786\\\hline
        2 (First points oscillate  &   NGC0300, IC4202, NGC0289, NGC0891, NGC2683, NGC2955, \\ with some intensity)
         & NGC3992, NGC4013, NGC4217, NGC5005, NGC5033, NGC5055,\\
         & NGC5371, NGC5907, NGC6195, NGC6946, NGC7814, PGC51017,\\
        & UGC02487, UGC02916, UGC02953, UGC03205, UGC03546, UGC03580,\\
        & UGC05253, UGC06614, UGC06786, UGC06973, UGC08699, UGC09992,\\
        & UGC11914 \\ \hline
        3 (Not enough data around & IC4202, NGC0289, NGC0891, NGC3726, NGC3992, NGC4051,\\ center of galaxy) 
        & NGC4138, NGC4217, NGC5005, NGC5033, NGC5055, NGC5371,\\
        & UGC02885, UGC06787, UGC09133 \\ 
        
    \hline
    \end{tabular}
    \caption{List of galaxies that are rejected in Sections~\ref{seccionpotencial} and~\ref{secciondensidad}, possibly due to defficiencies in the extraction of the part of the rotation curve $V_{DM}(r)$ assignable to dark matter. They are broken into specific behaviour classes as described in the text. Notice that some galaxies are present in  more than one of the classes.}
    \label{galaxias_descartadas}
\end{table}

\twocolumngrid



\end{document}